\documentclass{article}
\def \pul {{{\footnotesize{\frac{1}{2}}}}}

\def \BE {\begin{equation}}
\def \EE {\end{equation}}
\def \BEAH {\begin{eqnarray*}}
\def \EEAH {\end{eqnarray*}}
\def \BEA {\begin{eqnarray}}
\def \EEA {\end{eqnarray}}
\def \BDM {\begin{displaymath}}
\def \EDM {\end{displaymath}}

\def \T {\bigtriangleup}

\def \nt {\tau}

\def \pul {{{\scriptstyle{\frac{1}{2}}}}}

\def \eq#1 {\eqno{(#1)}}

\def \e#1 {{\rm e}^{#1}}

\def \ha { { 1\over 2 }}

\def \del {\partial}

\def \BE {\begin{equation}}
\def \EE {\end{equation}}

\def \bl {\mbox{\boldmath{$\ell$}}}

\def \bn {\mbox{\boldmath{$n$}}}

\def \hbm #1 {\mbox{\boldmath{$\hat m^{(#1)}$}}}
\def \bm {\mbox{\boldmath{$m$}}}

\def \cO {{\cal O}}

\def \mL {\mbox{\boldmath{$L$}}}
\def \mA {\mbox{\boldmath{$A$}}}
\def \mS {\mbox{\boldmath{$S$}}}

\def \mPsi {\mathbf \Psi}

\newcommand{\reals}{\mathbf{R}}

\newcommand{\fvec}[4]{{#1}_{\{a}{#2}_b{#3}_c{#4}_{d\}}}
\newcommand{\nlnm}[1]{\fvec{n}{\ell}{n}{m^{#1}}}
\newcommand{\nmnm}[2]{\fvec{n}{m^{#1}}{n}{m^{#2}}}
\newcommand{\nllm}[1]{\fvec{n}{\ell}{\,\ell}{m^{#1}}}
\newcommand{\lnlm}[1]{\fvec{\ell}{n}{\ell}{m^{#1}}}
\newcommand{\lmlm}[2]{\fvec{\ell}{m^{#1}}{\,\ell}{m^{#2}}}
\newcommand{\nmlm}[2]{\fvec{n}{m^{#1}}{\ell}{m^{#2}}}
\newcommand{\nlnl}{\fvec{n}{\ell}{n}{\ell}}
\newcommand{\nlmm}[2]{\fvec{n}{\ell}{m^{#1}}{m^{#2}}}
\newcommand{\nmmm}[3]{\fvec{n}{m^{#1}}{m^{#2}}{m^{#3}}}
\newcommand{\lmmm}[3]{\fvec{\ell}{m^{#1}}{m^{#2}}{m^{#3}}}
\newcommand{\mmlm}[3]{\fvec{m^{#1}}{m^{#2}}{\,\ell}{m^{#3}}}
\newcommand{\mmmm}[4]{\fvec{m^{#1}}{m^{#2}}{m^{#3}}{m^{#4}}}

\newcommand{\msub}[2]{m^{#1}{}_{#2}}
\newcommand{\msup}[2]{m_{#1}{}^{#2}}

\def\addots{\mathinner{\mkern1mu\raise1pt
    \vbox{\kern7pt\hbox{.}}\mkern2mu
    \raise4pt\hbox{.}\mkern2mu\raise7pt\hbox{.}\mkern1mu}}

\newcommand{\openbox}{\leavevmode
  \hbox to.77778em{%
  \hfil\vrule
  \vbox to.675em{\hrule width.6em\vfil\hrule}%
  \vrule\hfil}}

\newenvironment{proof}{\par\noindent{\bf Proof.}}{\openbox\par\medskip}
\newenvironment{proofof}[1]{\par\noindent{\bf Proof of
    #1.}}{\openbox\par\medskip}

\newcommand{\eqref}[1]{(\ref{#1})}
\newcommand{\tfrac}[2]{{\textstyle\frac{#1}{#2}}}
\newcommand{\supth}{^{\mathrm{th}}}

\newcommand{\VSIZ}{$\mbox{VSI}_0$ }

\newtheorem{proposition}{Proposition}
\newtheorem{theorem}[proposition]{Theorem}
\newtheorem{corollary}[proposition]{Corollary}
\newtheorem{lemma}[proposition]{Lemma}
\newtheorem{definition}[proposition]{Definition}

\newcommand{\lp}{\left(}
\newcommand{\rp}{\right)}

\newcommand{\rA}{\mathrm{A}}
\newcommand{\rB}{\mathrm{B}}
\newcommand{\rC}{\mathrm{C}}

\newcommand{\tC}{\mathbf{C}}

\newcommand{\tT}{\mathbf{T}}

\def \bk {\mbox{\boldmath{$k$}}}

\newcommand{\bo}{{\mathcal{B}}}

\def \bo {{\cal{B}}}

\newcommand{\SO}{\mathrm{SO}}

\newcommand{\sA}{{\scriptscriptstyle \rA}}

\def\overset#1#2{\mathord{\mathop{\kern0pt#2}\limits^{#1}}}
\newcommand{\ov}[2]{\overset{\,#2}{#1}}

\newcommand{\ovd}[3]{\overset{\,#2}{#1}{}_{#3}}
\newcommand{\ovu}[3]{\overset{\,#2}{#1}{}^{#3}}

\newcommand{\ovG}[1]{\overset{\;#1}{G}}

\newcommand{\ovQ}[1]{\overset{\,#1}{Q}}
\newcommand{\ovM}[1]{\overset{\,#1}{M}}
\newcommand{\Mi}{\ovM{i}}
\newcommand{\Mj}{\ovM{j}}
\newcommand{\Mk}{\ovM{k}}

\newcommand{\Xsup}[2]{X_{#2}{}^{#1}}

\newcommand{\Ysup}[2]{Y_{#2}{}^{#1}}

\newcommand{\ovduK}[3]{\ov{K}{#1}_{\!#2}{}^{#3}}

\newcommand{\ovdK}[2]{\ovd{K}{#1}{#2}}
\newcommand{\ovuK}[2]{\ovu{K}{#1}{#2}}

\newcommand{\ovdM}[2]{\ovd{M}{#1}{#2}}
\newcommand{\ovdE}[2]{\ovd{E}{#1}{#2}}

\newcommand{\ovdF}[2]{\ovd{F}{#1}{#2}}

\newcommand{\supp}[2]{{#1}^{(#2)}}
\newcommand{\suppsub}[3]{{#1}^{(#2)}\!{}_{#3}}

\begin{document}

\begin{center}
\Large {\bf Vanishing Scalar Invariant Spacetimes in Higher Dimensions}
\end{center}
\vspace{.3in}
\normalsize
\begin{center}
{\sc A. Coley$^1$, R. Milson$^1$, V. Pravda$^2$ and  A. Pravdov\' a$^2$} \\
\end{center}
\normalsize
\vspace{.6cm}
\begin{center}
{\em $^1$ Department of Mathematics and Statistics, Dalhousie University\\
Halifax, Nova Scotia B3H 3J5 \enskip Canada}\\
{\em $^2$ Mathematical Institute, Academy of Sciences, \v Zitn\' a 25,  115 67 Prague 1}
\vspace{0.3in}
\end{center}
\vspace{1cm}
%%%%%%%%%%%%%%%%%%%%%%%%%%%%%%%%%%%%%%%%%%%%%%%%%%%%%%%%%%%%%%%

\begin{abstract}

  We study manifolds with
  Lorentzian signature and prove that all scalar curvature invariants
  of all orders vanish in a higher-dimensional Lorentzian spacetime if
  and only if there exists an aligned non-expanding, non-twisting,
  geodesic null direction along
  which the Riemann tensor has negative boost order.
\end{abstract}

\section{Introduction}

Recently \cite{VSI} it was proven that in four-dimensional (4D)
pseudo-Riemannian or Lorentzian spacetimes all of the~scalar
invariants constructed from the~Riemann tensor and its covariant
derivatives of arbitrary order are zero if and only if
the~spacetime is of Petrov type III, N or O, all eigenvalues of
the~Ricci tensor are zero (the~Ricci tensor is consequently of
Pleba\' nski-Petrov type (PP-type) N or O, or alternatively, of
Segre type $\{ (31)\}$, $\{ (211)\}$ or $\{ (1111)\}$) and
the~common multiple null eigenvector of the~Weyl and Ricci tensors
is geodesic, shear-free, non-expanding, and non-twisting; we shall
refer to these spacetimes as vanishing scalar invariant (VSI)
spacetimes. An equivalent characterization of VSI spacetimes in 4D
is that there exists an aligned shear-free, non-expanding, non-twisting,
geodesic null direction $\ell^a$ along which the Riemann tensor
has negative boost order.  All of these spacetimes belong
to Kundt's class, and hence the~metric of these spacetimes can be
expressed in an appropriate form in adapted coordinates
\cite{kramer,kundt}.  VSI spacetimes can be classified according
to their Petrov type, Segre type and the vanishing or
non-vanishing of the quantity $\tau$.  This leads to 16 non-trivial
distinct classes of VSI spacetimes, one of which is the vacuum
pp-wave (Petrov type N, vacuum, $\tau=0$) spacetime, in additional
to the trivial flat Minkowski spacetime. All of the corresponding
metrics are displayed in \cite{VSI}. The generalized pp-wave
solutions are of Petrov-type N, PP-type O (with $\nt=0$), and
admit a covariantly constant null vector field \cite{jordan}.

We shall study VSI spacetimes in arbitrary $N$ dimensions (not necessarily even,
but $N=10$ is of particular importance from string theory) and, in principle,
for arbitrary signature. However, we shall focus our attention on Lorentzian
manifolds with signature $N-2$. We note that for
Riemannian manifolds with signature $N$, flat space is the only VSI manifold.
Manifolds with signature $N-4$ with $N \ge 5$ are also of physical interest \cite{ss,wesson}.
In \cite{cmpppz} we investigated $N$-dimensional Lorentzian spacetimes
in which all of the~scalar invariants
constructed from the~Riemann tensor and its
covariant derivatives are
zero. These spacetimes are  higher-dimensional
generalizations of $N$-dimensional pp-wave spacetimes,
which have been of interest recently in the context of string theory
in curved backgrounds in higher dimensions.
We presented a canonical form for the Riemann and Weyl tensors
in a preferred null frame in arbitrary dimensions if all of the scalar curvature invariants vanish
(thereby generalizing the theorem of \cite{VSI} to higher dimensions).
We shall prove the assertions in this paper, and we shall briefly discuss the algebraic
structure of the resulting spacetimes. In particular, we shall prove:

\begin{theorem}
\label{main-theorem}

All curvature invariants of all orders vanish in an
$N$-dimensional Lorentzian spacetime if and only if there exists
an aligned non-expanding ($S_{ij}=0$), non-twisting ($A_{ij}=0$),  geodesic null direction
$\ell^a$ along which the Riemann tensor has negative boost order.

\end{theorem}

An analytical form of the conditions in Theorem \ref{main-theorem}
are as follows: \BE R_{abcd} = 8 A_{i} \, \lnlm{i}+ 8 B_{ijk} \,
\mmlm{i}{j}{k}+ 8 C_{ij} \, \lmlm{i}{j} \label{RRR} \EE (i.e., the
Riemann tensor is of algebraic type III or N \cite{newcqg}), and \BE \ell_{a
; b} = L_{11} \ell_a \ell_b  + L_{1i} \ell_a \msub{i}{b} + L_{i1}
\msub{i}{a} \ell_b; \label{l_ab-VSI} \EE 
that is, the expansion
matrix $S_{ij}=0$, the twist matrix $A_{ij}=0$ (which are the
analogues of $\rho$, $\sigma$ in 4D; see Section 1.2 for the
definitions), as well as $L_{i0}=0=L_{10}$
(corresponding to an affinely parametrized geodesic congruence $\ell_a$;
i.e., analogues of $\kappa$, $\epsilon + \bar{\epsilon}$, respectively -- see Eq.
(\ref{dl})).

In this paper we shall first summarize the N dimensional null
frame formalism, the algebraical classification of a tensor based
on boost order \cite{cmpppz,Algclass,newcqg} and the Bianchi
identities and their consequnces for vacuum type III and N
spacetimes \cite{Bianchi}. The sufficiency and necessity of
Theorem \ref{main-theorem} are then proven in Sections
\ref{sec-suff} and \ref{sec-nec0}, \ref{sec-nec}, respectively.
The paper concludes with a discussion. Many of the details of the
analysis are found in the Appendices.

\subsubsection{Notation}

We shall consider a null frame   $\bl=\bm_0,\ \bn=\bm_1,\ \bm_2,
...\ \bm_{N-1}$ ($\bl,\ \bn$ null with  $\ell^a \ell_a= n^a n_a =
0$, $\ell^a n_a = 1$, $\bm_i$ real and spacelike $m_i{}^a m_j{}_a
= \delta_{ij}$, $i=2,...,N-1$,  all other products vanish) in an
$N$-dimensional Lorentz-signature space(time), so that \BE g_{ab}
= 2\ell_{(a}n_{b)} +  \delta_{jk} m^j{}_a m^k{}_b. \label{tetrad}
\EE Covariance is relative to the group of linear Lorentz
transformations.  Throughout, Roman indices $a,b,c, \rA, \rB, \rC$
range from $0$ to $N-1$.  Lower case indices indicate an arbitrary
basis, while the upper-case ones indicate a null frame. Space-like
indices $i,j,k$ also indicate a null-frame, but vary from $2$ to
$N-1$ only.   We will raise and lower the space-like indices using
$\delta_{ij}$; e.g., $T_i =  \delta_{ij}T^j$. We will observe
Einstein's summation convention for both  of these types of
indices; however, for indices $i,j \dots$ there is no difference
between covariant and contravariant components and thus we will
not distinguish between subscripts and superscripts.

We also introduce the notation (compare with \cite{cmpppz,newcqg})
\BE w_{\{a} x_b y_c z_{d\}} \equiv \frac{1}{2}(w_{[a} x_{b]}
y_{[c} z_{d]}+
 w_{[c} x_{d]} y_{[a} z_{b]}) \equiv \frac{1}{8}\{[w_p x_q] [y_r z_s]\} .\label{zavorka}
\EE

\subsection{Background}

A {\em null rotation} about $\bn$ is a
Lorentz transformation of the form
\begin{equation}
  \label{eq:nullrot}
    \hat{\bn}=  \bn,\quad
    \hat{\bm}_i=  \bm_i + z_i \bn,\quad
    \hat{\bl}= \textstyle \bl
    -z_i \bm^i-\frac{1}{2} \delta^{ij} z_i z_j\, \bn.
\end{equation}
A null rotation about $\bl$ has an analogous form.  A boost is a
transformation of the form
\begin{equation}
  \label{eq:boost}
    \hat{\bn}= \lambda^{-1}\bn,\quad
    \hat{\bm}_i=  \bm_i,\quad
    \hat{\bl}=  \lambda\, \bl, \quad \lambda \neq 0.
\end{equation}
A spin is a transformation of the form
\begin{equation}
  \label{eq:spin}
    \hat{\bn}= \bn, \quad
    \hat{\bm}_i=  X_i^j\,\bm_j,\quad
    \hat{\bl}=  \bl,
\end{equation}
where $X_i^j$ is an orthogonal matrix.

Let $T_{a_1... a_p}$ be a rank $p$ tensor.  For a fixed list of
indices $A_1,...,A_p$, we call the corresponding $T_{A_1...
  A_p}$ a null-frame scalar.  These scalars transform under a boost
(\ref{eq:boost}) according to
\begin{equation}
  \label{eq:boostxform}
  \hat{T}_{A_1... A_p}= \lambda^b\,
  T_{A_1... A_p},\quad b=b_{A_1}+...+b_{A_p},
\end{equation}
where
\begin{equation}
  \label{eq:bwvalues}
  b_0=1,\quad b_i=0,\quad b_1=-1.
\end{equation}
We call the
above $b$ the {\em boost-weight} of the scalar.
We define the {\em boost order} of the tensor $\tT$, as a whole, to be
the boost weight of its leading term \cite{Algclass}.

We can then decompose the Riemann tensor and sort its components 
by boost weight:
\BEA
\label{eq:riemexp}
  R_{abcd} &=&
  \overbrace{
    4 R_{0i0j}\, \nmnm{i}{j}}^2  +\overbrace{
    8R_{010i}\, \nlnm{i} +
    4R_{0ijk}\, \nmmm{i}{j}{k}}^1 \\
  \nonumber
  &&+\left\{
    \begin{array}{l}
      +4 R_{0101}\, \nlnl +  4 R_{01ij}\, \nlmm{i}{j} \\
      +8 R_{0i1j}\, \nmlm{i}{j} +  R_{ijkl}\, \mmmm{i}{j}{k}{l}
    \end{array}
  \right\}^0 \\
  \nonumber &&+ \overbrace{ 8 R_{101i}\, \lnlm{i}+
    4 R_{1ijk}\, \lmmm{i}{j}{k}}^{-1} +
  \overbrace{ 4 R_{1i1j}\, \lmlm{i}{j} }^{-2}.  \EEA
 The Weyl tensor $C_{abcd}$ has a
  boost-weight decomposition analagous to \eqref{eq:riemexp}.  Table
  \ref{tab:bw2} shows the boost weights for the scalars of the Weyl
  curvature tensor $C_{abcd}$.  Thus, generically $C_{abcd}$ has boost
  order $2$.  If all $C_{0i0j}$ vanish, but some $C_{010i}$, or
  $C_{0ijk}$ do not, then the boost order is $1$, etc.  The Weyl
  scalars also satisfy a number of additional relations, which follow
  from curvature tensor symmetries and from the trace-free condition:
\begin{eqnarray}
  C_{0i0}{}^i = 0, \nonumber \\
  C_{010j} = C_{0ij}{}^i,  C_{0(ijk)} = 0,\nonumber\\
  C_{0101} = C_{0i1}{}^i,  C_{0i1j}=-\frac{1}{2} C_{ikj}{}^k+\frac{1}{2}C_{01ij},
  C_{i(jkl)} = 0,\nonumber\\
  C_{011j} = -C_{1ij}{}^i,  C_{1(ijk)} = 0,\nonumber\\
  C_{1i1}{}^i = 0.
\end{eqnarray}

\begin{table}[htbp]
  \begin{center}
    \begin{tabular}{|c|c|c|c|c|}
      \hline
      $2$ & $1$ & $0$ & $-1$ & $-2$\\
      \hline
      $C_{0i0j}$ &  $C_{010i}, C_{0ijk}$ & $C_{0101}, C_{01ij},
      C_{0i1j}, C_{ijkl}$ & $C_{011i}, C_{1ijk}$ & $C_{1i1j}$\\
      \hline
    \end{tabular}
    \medskip
    \caption{Boost weights of  the Weyl scalars.}
    \label{tab:bw2}
  \end{center}
\end{table}

A priori, the assignment of a boost order to a tensor seems to depend
on the choice of a null-frame  \cite{Algclass}.  However, a null rotation about $\bl$
fixes the leading terms of a tensor, while boosts and spins subject
the leading terms to an invertible transformation.  It follows that
the boost order of a tensor is a function of the null direction $\bk$.
We shall therefore denote boost order by $\bo(\bk)$ (with the choice
of a tensor $\tC$ determined by context).
We will say that a null vector $\bk$ is {\em aligned} with the Weyl
tensor $\tC$ whenever ${\bo(\bk)}\leq 1$  \cite{Algclass,newcqg}.  We will call the integer
$1-{\bo(\bk)}\in \{0,1,2,3\}$ the {\em order of alignment}.

\begin{definition}
 We will say that the {\bf principal type} of a Lorentzian manifold
  is I, II, III, N according to whether there exists an aligned $\bk$
  of alignment order $0,1,2,3$, respectively (i.e.,
  ${\bo(\bk)}=1,0,-1,-2$, respectively).  If no aligned $\bk$
  exists we will say that the manifold is of type G.  If the Weyl
  tensor vanishes, we will say that the manifold is of type O.
\end{definition}

It follows that there exists a frame in which the components of the
Weyl tensor satisfies:
\begin{eqnarray}
&&  Type\ I:\ \ ~~ C_{0i0j}=0, \nonumber\\
&&  Type\ II: \ ~~C_{0i0j}=C_{0ijk}=0, \nonumber \\
 && Type\ III:  ~~C_{0i0j}=C_{0ijk}=C_{ijkl} =C_{01ij}=0, \nonumber\\
 && Type\ N:  \ ~~C_{0i0j}=C_{0ijk}=C_{ijkl} = C_{01ij}=C_{1ijk}=0.
\end{eqnarray}
The general types have various algebraically special subtypes
\cite{newcqg}, which include (the following only lists the
additional conditions for the algebraic specializations): $Type\
Ia  ~~C_{010i}=0$,  subclasses of $Type\ II$ (with $C_{0101}=0$, the
traceless Ricci part of $C_{ijkl} = 0$, the Weyl part of $C_{ijkl}
= 0$, and $C_{01ij} = 0$), and $Type\ IIIa ~~ C_{011i}=0$. Note
that the conditions for the  type II subcases can be combined to
yield composite types.  The {\bf full type} of
the Weyl tensor includes identifying its principal type and
subclass and multiplicities, etc. \cite{Algclass}. The special type D
is defined by the fact that in canonical form all terms are of boost weight zero.
For type III
tensors, the principle null direction (PND) of order 2 is unique.
There are no PNDs of order 1, and at most 1 PND of order 0.  (For
$N=4$ there is always exactly $1$ PND of order 0; for $N>4$ this
PND need not exist.) For type N tensors, the order 3 PND is the
only PND of any order.

\subsection{Bianchi Identities}

Covariant derivatives of the frame vectors can be expressed as
\cite{Bianchi}: \BEA \ell_{a ; b} &=& L_{11} \ell_a \ell_b + L_{10}
\ell_a n_b + L_{1i} \ell_a \msub{i}{b} + L_{i1} \msub{i}{a} \ell_b +
L_{i0} \msub{i}{a} n_{b} + L_{ij}
\msub{i}{a} \msub{j}{b}  , \label{dl} \\
n_{a ; b } &=& -L_{11} n_a \ell_b -L_{10} n_{a } n_{b }-L_{1i} n_a
\msub{i}{b} + N_{i1} \msub{i}{a} \ell_b + N_{i0} \msub{i}{a} n_b +
N_{ij} \msub{i}{a}
\msub{j}{b}  , \label{dn} \\
\msub{i}{a;b} &=& -N_{i1} \ell_a \ell_b -N_{i0} \ell_a n_b -{L}_{i1}
n_a \ell_b
-L_{i0} n_a n_b -N_{ij} \ell_a \msub{j}{b}   \nonumber \\
&&+ {\Mi}_{j1} \, \msub{j}{a} \ell_b -L_{ij} n_a \msub{j}{b} +
{\Mi}_{j0}\, \msub{j}{a} n_b + {\Mi}_{kl}\, \msub{k}{a} \msub{l}{b} 
\label{dm} \EEA
(where $L_{11}$ is the analogue of the spin coefficient $\gamma +\bar{\gamma}$ in 4D, etc.).

Let us decompose ${\mL}$ into its symmetric and antisymmetric parts,
${\mS}$ and  ${\mA}$,
\BE
L_{ij}=S_{ij} + A_{ij}, \quad S_{ij} = S_{ji} , \ A_{ij}= -A_{ji}.\label{Lij}
\EE
If $\bl$ corresponds to a null geodesic congruence with an affine parametrization,
we can express the expansion $\theta$ %, twist $\omega$
and the shear matrix $\sigma_{ij}$ as
\BEA
\theta &\equiv &\frac{1}{n-2} \ell^{a}_{\ ; a} = \frac{1}{n-2}[\mS] ,\\
\sigma_{ij} &\equiv& \left(
\ell_{(a ; b)}
- \theta \delta_{kl}   \,\msub{k}{a} \msub{l}{b} \right)
\msup{i}{a} \msup{j}{b}
=S_{ij}-\frac{[\mS]}{n-2}\delta_{ij}. \label{shearmatrix}
\EEA
%We will also denote $\sigma^2\equiv\sigma_{ij} \sigma_{ij}$.
For simplicity, let us call $\mA$ the {\it twist matrix} and $\mS$  the
{\it expansion matrix}, although
$\mS$ contains information about both expansion and shear.
We also introduce the quantities
\BE
S\equiv\pul[\mS],\ \ A^2\equiv\pul A_{ij}A_{ij}.
\EE

%\BEA
%\theta &\equiv &\frac{1}{2} \ell^{a}_{\ ; a} = \frac{1}{2}[S] ,\\
%\omega^2 &\equiv& \frac{1}{2} \ell_{[a ; b]} \ell^{a ; b} = \frac{1}{2}A_{ij} A_{ij} ,\\
%|\sigma|^2 &\equiv& \frac{1}{2} \ell_{(a ; b)} \ell^{a ; b}- \frac{1}{4} (\ell^{a}_{\ ;a})^2 =
%\frac{S_{ij} S_{ij}}{2} - \frac{[S]^2}{4} .
%\EEA
%We will thus call the matrix $S_{ij}$  {\it expansion matrix} and  $A_{ij}$  {\it twist matrix}.

We next introduce directional derivatives $D$, $\T$, and
$\delta_i$ by \BE D \equiv \ell^a \nabla_a, \ \ \ \bigtriangleup
\equiv n^a \nabla_a, \ \ \ \delta_i \equiv \msup{i}{a} \nabla_a ,
\quad \nabla_a = n_a D + \ell_a \T + \msub{i}{a} \delta_i.
\label{covder} \EE Commutators then have the form \BEA
\T D - D \T &=& L_{11} D + L_{10} \T + L_{i1} \delta_i - N_{i0} \delta_i, \\
\delta_i D - D \delta_i &=& (L_{1i} + N_{i0}) D + L_{i0} \T + (L_{ji}-\Mi_{j0}) \delta_j, \\
\delta_i \T - \T \delta_i  &=& N_{i1} D + (L_{i1}-L_{1i}) \T + (N_{ji}-\Mi_{j1}) \delta_j, \\
\delta_i \delta_j - \delta_j \delta_i &=& (N_{ij}-N_{ji}) D + (L_{ij}-L_{ji})\T
 + (\Mj_{ki}-\Mi_{kj})\delta_k.
\EEA

For type III and type  N vacuum spacetimes  we will use the
notation of \cite{Bianchi} with \BDM \Psi_{i} = C_{101i}, \ \
\Psi_{ijk}= \frac{1}{2} C_{1kij}, \ \ \Psi_{ij} = \frac{1}{2}
C_{1i1j} . \EDM The Weyl tensor can be thus expressed as \BE
C_{abcd} = 8 \Psi_{i} \lnlm{i} + 8 \Psi_{ijk} \mmlm{i}{j}{k}  + 8
\Psi_{ij} \lmlm{i}{j} . \label{WeylIIIN} \EE Case $\Psi_{ijk}
\not= 0$ is of the type III, while $\Psi_{ijk} = 0$ (and
consequently also $\Psi_{i} = 0$) corresponds to type N.
Note that $\Psi_{ij}$ is symmetric and traceless. $\Psi_{ijk}$ is antisymmetric in the first two indices
and in vacuum also satisfies{\footnote{Note that we use two
different  operations denoted by $\{\}$. In the first case $\{\}$ act on three indices
and stands for $R_{ab\{cd;e\}}=R_{abcd;e}+R_{abde;c}+R_{abec;d}$.
In the other case $\{\}$ act on four indices and is given by (\ref{zavorka}). }
}
\BEA
\Psi_i=2 \Psi_{ijj}, \label{eqPsi_i}\\
   \Psi_{\{ijk\}}=0. \label{eqPsicykl}
\EEA

Further constraints on $\Psi_{ij}$, $\Psi_{ijk}$, $\mS$, $\mA$ and $\bl$ can be obtained by
employing the Bianchi and Ricci identities
\BEA
&&R_{ab\{cd;e\}}=0 \label{Bia}, \\
&&V_{a;bc}=V_{a;cb}+R^s_{\ abc} V_s,
\EEA
where ${\mbox{\boldmath{$V$}}} $ is an arbitraty vector.

The implications of the Bianchi identities (\ref{Bia}) for vacuum type III and N spacetimes
were studied in \cite{Bianchi}.
It was shown that for these spacetimes $\bl$ is geodesic.
Furthermore, if they have non-vanishing expansion $\mS$ and twist $\mA$ and
${\mbox{\boldmath{$u$}}}^{\alpha}$ , {\mbox{\boldmath{$v$}}} , and {\mbox{\boldmath{$w$}}}
are orthonormal elements of the vector space spanned by vectors $\bm^{(i)}$ (see Sec. III and IV in \cite{Bianchi})
then:\\
 1) In vacuum type N spacetimes with $S \not=0$ it is always possible to choose vectors
{\mbox{\boldmath{$v$}}}  and {\mbox{\boldmath{$w$}}} for which
\BEA
\Psi_{ij}& =& \sqrt{\frac{p}{2}}  (v_i v_j - w_i w_j)  ,\label{PsitypeN}\\
S_{ij} &=& S (v_i v_j + w_i w_j),  \label{decompPsiS}\\
A_{kl}& =& {A} (w_k v_l  - v_k w_l) \label{decompA} ,
\EEA
where
\BE \label{eq:pdef} p\equiv\Psi_{ij}\Psi_{ij}.\EE

2)
In vacuum type III spacetimes with $S \not=0,\ A\not=0,\ \Psi_i \not=0$
and with ``general form'' of $\Psi_{ijk}$ (see \cite{Bianchi} for details)
 it is possible to introduce a vector
\BE
\Phi_i \equiv A_{ij} \Psi_j
\EE
and then express $\mS$, $\mA$, and $\Psi_{ijk}$ as
\BEA
S_{ij} &=& S\left( \frac{\Psi_i \Psi_j}{\psi^2} +  \frac{\Phi_i \Phi_j}{\phi^2}      \right) ,\label{vyslS}\\
A_{ij} &=& \frac{1}{\psi^2} (\Phi_i \Psi_j - \Phi_j \Psi_i), \\
\Psi_{ijk}&=& \frac{1}{2\phi^2} (\Psi_i \Phi_j - \Psi_j \Phi_i) \Phi_k \nonumber \\
&+& \frac{\cO_{1\alpha 1}}{\psi^2} (\Psi_i u_j^{\alpha}-\Psi_j u_i^{\alpha}) \Psi_k
- \frac{\cO_{1\alpha 1}}{\phi^2} (\Phi_i u_j^{\alpha}-\Phi_j u_i^{\alpha}) \Phi_k \nonumber  \\
&+& \frac{\cO_{1\alpha 2}}{{\psi \phi}} (\Psi_i
u_j^{\alpha}-\Psi_j u_i^{\alpha}) \Phi_k + \frac{\cO_{1\alpha
2}}{{\psi \phi}} (\Phi_i u_j^{\alpha}-\Phi_j u_i^{\alpha}) \Psi_k
,  \label{vyslB} \EEA where $\phi^2 \equiv \Phi_i \Phi_i$,
$\psi^2=\Psi_i \Psi_i$. We have treated all possible degenerate
situations for the non-twisting case, $A_{ij}=0$, in  arbitrary
dimensions (see App. C.1 in \cite{Bianchi}) and for the twisting
case in five dimensions (see App. C.2 in \cite{Bianchi}), and they
all lead to special cases of the solution
(\ref{vyslS})--(\ref{vyslB}) which are given explicitly in
\cite{Bianchi}. We assert that the solution
(\ref{vyslS})--(\ref{vyslB}) is a general solution even for the
twisting case in arbitrary dimensions. We have proven this in all
of the general cases, and although we have not rigorously proven
this for all of the degenerate cases there is evidence that it is
indeed true.

\section{The Sufficiency Proof}
\label{sec-suff}

In this section we start with the assumptions of Theorem
\ref{main-theorem}  and  then show that all curvature invariants
of all orders vanish. The corresponding form of the Riemann
tensor (\ref{RRR}), which implies appropriate types for the Weyl and Ricci
tensors, was given in \cite{cmpppz}. The congruence
corresponding to $\bl$ is geodesic with vanishing  shear, twist
and expansion; i.e., \BE L_{i0}=L_{ij}=0. \EE Let us, for
simplicity, choose an affine parametrization and a parallely
propagated frame. Thus  \BEA
L_{10}=0, \ \ %n_{a;b} \ell^b = 0 \Rightarrow
N_{i0}=0, \ \ %\\ m^{(i)}_{a;b} \ell^b = 0 \Rightarrow
\Mi_{j0}=0. \EEA Due to the Bianchi identities \BE
R_{ab\{cd;e\}}=0, \label{Bianchi} \EE we can express how the
operator $D$ acts on $A_i$, $B_{ijk}$, and  $C_{ij}$.
Corresponding results may be obtained by evaluating  the Bianchi
identities with various combinations of  null-frame indices, and
by using the form (\ref{RRR}) of the Riemann tensor. For example,
for indices $101i0$ we obtain \BE D A_i=0. \label{BianchiAi} \EE
%means that the equation $D A_i=0$ can be obtained by contracting
%(\ref{Bianchi}) with $n^a \ell^b n^c m^{(i)d} \ell^e$.
The other two equations of interest are obtained by using,
 respectively, indices $ij1k0$ and $1i1j0$:
\BEA &&D B_{ijk}=0, \label{BianchiBijk} \\
  &&D C_{ij} = -B_{kji} L_{k1} + A_i L_{[1j]} + \frac{1}{2}
\Mk_{ij} A_k + \frac{1}{2} \delta_j A_i.  \label{BianchiCij} \EEA
Evaluating the Ricci identities
\BE
\ell_{a; b c } - \ell_{a; c b} = R^{d}_{\ a b c} \ell_d
 \label{Ricl}
\EE and using indices, $10i$, $i01$, and $110$, respectively,
yields \BEA
&&D L_{1i} = 0,  \label{RicciL1i} \\
&&D L_{i1} = 0, \label{RicciLi1}  \\
&&D L_{11} = -L_{1i} L_{i1} \label{RicciL11}.
\EEA
From Eqs. (\ref{RicciL1i})--(\ref{RicciL11}), it follows that
\BE
D^2 L_{11} =0.
\EE
Similarly, evaluating
\BE
n_{a; b c } - n_{a; c b} = R^{d}_{\ a b c} n_d  \label{Ricn}
\EE
with indices $ij0$ and $i10$, respectively, we get
\BEA
&& D N_{ij} = 0, \label{RicNij} \\
&& D N_{i1} = -N_{ij} L_{j1} + A_i. \label{RicNi1}  \EEA From Eqs.
(\ref{BianchiAi}), (\ref{RicciLi1}), and (\ref{RicNij}) we obtain
\BE D^2 N_{i1}=0. \EE Evaluating \BE \msub{i}{a; b c } -
\msub{i}{a; c b} = R^{d}_{\ a b c} \msub{i}{d} \label{Ricm} \EE
with indices $j0k$ and $j10$, respectively, we obtain\BEA
&& D  \Mi_{jk}  = 0, \label{RicciMijk} \\
&& D  \Mi_{j1}  = - \Mi_{jk} L_{k1}.
\EEA
From Eqs.
(\ref{RicciLi1}) and (\ref{RicciMijk}) we get
\BE
D^2 \Mi_{j1}  =0.
\EE
Note that from the previous equations it also follows that
\BE
D^2 C_{ij} = 0.
\EE

\begin{table}
\begin{center}

\begin{tabular}{lcl}
\hline \hline
quantity & boost weight & "D-equation" \\
\hline \hline \\
$L_{11}%=\ell_{a ; b } n^{a } n^{b }
$ & -1 & $D^2 L_{11} = 0$ \\
$L_{1i}%=-\ell_{a ; b } n^{a } m^{(i)b }
$ & 0 & $D L_{1i} = 0$ \\
$L_{i1}%=-\ell_{a ; b } m^{(i)a } n^{b }
$ & 0 & $D L_{i1} = 0$ \\
\hline
%$N_{21}=n_{a ; b } \ell^{a } n^{b }$ & -1 & $D^2 N_{21} = 0$ \\
$N_{i1}%=-n_{a ; b } m^{(i) a } n^{b }
$ & -2 & $D^2 N_{i1} = 0$ \\
%$N_{2i}=-n_{a ; b } \ell^{a } m^{(i) b }$ & 0 & $D N_{2i} = 0$ \\
$N_{ij}%=n_{a ; b } m^{(i) a } m^{(j) b }
$ & -1 & $D N_{ij} = 0$ \\
\hline
$\Mi_{jk}%=m^{(i)}_{a ; b } m^{(j) a } m^{(k) b }
$ & 0 & $D \Mi_{jk} = 0$ \\
$\Mi_{j1}%=-m^{(i)}_{a ; b } m^{(j) a } n^{ b }
$ & -1 & $D^2 \Mi_{j1} = 0$ \\
%$\Mi_{1j}=-m^{(i)}_{a ; b } n^{ a } m^{(j) b }$ & -1 & $D \Mi_{1j} = 0$ \\
%$\Mi_{21}=m^{(i)}_{a ; b } \ell^{ a } n^{ b }$ & 0 & $D \Mi_{21} = 0$ \\
%$\Mi_{11}=m^{(i)}_{a ; b } n^{ a } n^{ b }$ & -2 & $D^2 \Mi_{11} = 0$ \\
\hline
$A_{i}$ & -1 & $D A_i=0$ \\
$B_{ijk}$ &-1& $DB_{ijk}=0$\\
$C_{ij}$ &-2& $D^2 C_{ij}=0$\\
\hline \hline
\end{tabular}
\caption{\label{tableboostD} Properties of quantities which are relevant for our proof.}
\end{center}
\end{table}

Let us now express covariant derivatives of the frame vectors and
commutators for a geodesic, affinely parametrized, expansion and
twist free $\bl$ and the rest of the frame parallely propagated
along $\bl$ \BEA \ell_{a ; b } &=& L_{11} \ell_a \ell_b +L_{1i}
\ell_a \msub{i}{b}  +
L_{i1} \msub{i}{a} \ell_b  \label{dl} ,\\
n_{a ; b } &=& -L_{11} n_a \ell_b -L_{1i} n_a \msub{i}{b}  +
N_{i1} \msub{i}{a} \ell_b  + N_{ij} \msub{i}{a} \msub{j}{b} \label{dn} ,\\
\msub{i}{a ; b } &=& -{N}_{i1} \ell_a \ell_b-{L}_{i1} n_a \ell_b
-{N}_{ij} \ell_a \msub{j}{b}  +
{\Mi}_{j1} \msub{j}{a} \ell_b  + {\Mi}_{kl} \msub{k}{a} \msub{l}{b}
\label{dm} ;\\
%\EEA
 &&\nonumber\\
%Since all basis vectors are paralelly propagated along $\ell^{a}$,
%\BE
%D \ell^a = D n^a = D m^{(i) a} = 0.
%\EE
%\BEA
\T D - D \T &=& L_{11} D + L_{i1} \delta_i,  \label{comTD} \\
\delta_i D - D \delta_i &=& L_{1i} D , \label{comdD} \\
\delta_i \T - \T \delta_i  &=& N_{i1} D + (L_{i1}-L_{1i}) \T + (N_{ji}-\Mi_{j1}) \delta_j ,\\
\delta_i \delta_j - \delta_j \delta_i &=& (N_{ij}-N_{ji}) D %+ (L_{ij}-L_{ji})\T
 + (\Mj_{ki}-\Mi_{kj})\delta_k.
\EEA

Now we can proceed with the proof in a similar way to that in four
dimensions \cite{VSI}. Thus, we will  only outline the key points.
In all of the proofs   we recall that $\bl$ is geodesic and
affinely parametrized, the expansion and twist matrices vanish and
the frame is parallely propagated along $\bl$.

\begin{definition}
We shall say that a weighted scalar $\eta$ with boost weight $b$ is
balanced if $D^{-b} \eta =0$ for $b<0$ and $\eta=0$ for $b \geq 0$.
\end{definition}

In analogy with the proof in 4D
(employing Table \ref{tableboostD} and the commutators (\ref{comTD}), (\ref{comdD})) we obtain
%then leads to

\begin{lemma}
\label{lemmabal1}
If $\eta$ is a balanced scalar then \\
 \hspace*{3cm} $L_{11} \eta , \  L_{1i} \eta ,\ L_{i1} \eta ,\ N_{i1}
 \eta ,\ N_{ij} \eta ,\ \Mi_{\!j1}\,  \eta ,\
\Mi_{\!kl}\,  \eta, \
 D \eta ,\ \delta_i \eta,\ \T \eta $  \\
are balanced as well.
\end{lemma}

\begin{definition}
A balanced tensor is a tensor whose components are all balanced scalars.
\end{definition}

From (\ref{covder}) and Lemma \ref{lemmabal1}
(together with the assumptions in this section), it then follows that

\begin{lemma}
  A covariant derivative of an arbitrary order of a balanced tensor is
  again a balanced tensor.
\end{lemma}

The Riemann tensor in the form (\ref{RRR}) is balanced from Eqs.
(\ref{BianchiAi}), (\ref{BianchiBijk}), (57).
Consequently, all of its covariant derivatives are  also balanced and thus
all curvature invariants in this case vanish. This completes the
sufficiency part of the proof.

\section{The  Necessity Proof for Zeroth Order Invariants}
\label{sec-nec0}

In this section we prove that the Riemann tensor for VSI spacetimes necessarily has
negative boost order.

A scalar formed from the contractions of the Riemann curvature
tensor is an invariant quantity in the sense that the contraction
yields the same answer with respect to every choice of basis.
There is an infinite number of different ways to perform such
contractions. Hence, a given curvature tensor has associated with
it an infinite set of invariants, some of which are generically
non-vanishing.  In this section we classify the algebraically
special spacetimes characterized by the condition that all zeroth
order (i.e., algebraic) invariants formed from $R_{abcd}$ vanish.
Henceforth we shall refer to such spacetimes as belonging to the
\VSIZ class. We will show that such spacetimes are necessarily of
principal type III, N, or O, with an aligned Ricci tensor of type
PP-N, PP-O (or vacuum).

A scalar invariant has boost weight zero.  It follows immediately
that if the boost order of the curvature tensor is negative along
some aligned null direction $\ell^a$, then all scalar invariants
must vanish.  The converse is much harder to prove.  In 4D
it is well known \cite{kramer,penrind} that the
vanishing of the fundamental second and third-degree Weyl
invariants, $I=J=0$, implies that the principal type is III, N, or O. The
condition that the Ricci tensor either vanishes or is of type PP-N
or PP-O and is aligned, follows easily.

In higher dimensions, an entirely different approach is needed.
We define a curvature-like tensor to be a rank~4 tensor with the
following index
symmetries:
$$R_{abcd} = -R_{bacd} = R_{cdab}.$$
The class of curvature-like
tensors is more general than the class of Riemann curvature tensors,
because we do not impose the algebraic Bianchi condition
\begin{equation}
  \label{eq:rcond}
  R_{abcd}+ R_{acdb}+R_{adbc}=0.
\end{equation}
Indeed, curvature-like tensors may be best characterized as
symmetric, rank 2 tensors with bivector indices. On occasion we
will, therefore,  write curvature-like tensors as $R_{\alpha\beta}
= R_{\beta\alpha},$ where $\alpha, \beta$ are bivector indices as
defined in Appendix \ref{ap:bivectors}.

To every curvature-like $R_{abcd}$ we associate the rank 2,
symmetric covariant $$R_{ab} = R_{acb}{}^c,$$ which we will call
the Ricci covariant.  In addition, we could raise a bivector index
and consider the transformation of bivector space
$R_\alpha{}^{\!\beta}$ (see Appendix C).  By taking powers  and then lowering an
index we obtain additional covariants: e.g., the following
curvature like tensors $\supp{R}{k}_{\;\alpha\beta}$,
$$
\supp{R}{2}_{\;\alpha\beta} = R_\alpha{}^{\gamma}\, R_{\gamma\beta},\quad
\supp{R}{3}_{\;\alpha\beta} = R_\alpha{}^\gamma R_\gamma{}^\delta
R_{\delta\beta},\quad \mbox{etc.}
$$
% We also have the Ricci covariants of the powers,
% $$\supp{R}{k}{}_{ab} = \supp{R}{k}{}_{acb}{}^c.$$
Our main result is the following.
\begin{theorem}
  \label{th:VSIcform}
  Let $R_{abcd}$ be a non-zero, curvature-like tensor with vanishing
  zeroth order invariants.  Then, $\supp{R}{3}_{\;\alpha\beta} =0,$ and
  there exists an aligned null direction $\ell^a$ along which the boost
  order is negative.  Generically, that is for
  $\supp{R}{2}_{\;\alpha\beta} \neq0,$ the principal type is III, with
  \begin{equation}
    \label{eq:ptypeIII}
    R_{abcd} =8R_{011i}\, \nllm{i} + 4R_{1ijk}\, \lmmm{i}{j}{k}
    +4R_{1i1j}\,  \lmlm{i}{j}.
  \end{equation}
  If $\supp{R}{2}_{\;\alpha\beta}=0$, then the principal type is N, with
  \begin{equation}
    \label{eq:ptypeN}
    R_{abcd} = 4 R_{1i1j}\,  \lmlm{i}{j}.
  \end{equation}
\end{theorem}

The proof will be broken up into a number of lemmas that treat special
cases (the proofs of the lemmas are given in the Appendices).

\begin{lemma}
  \label{lem:pwzero}
  A curvature-like tensor of pure boost-weight zero possesses a
  non-vanishing invariant.
\end{lemma}

\begin{lemma}
  \label{lem:nzr}
  Let $R_{abcd}$ be a curvature-like tensor with vanishing zeroth
  order invariants.
  If the Ricci covariant $R_{ab}\neq 0$, then there exists an
  aligned null direction $\ell^a$ along which the boost order of
  $R_{abcd}$ is negative.
\end{lemma}

% \begin{lemma}
%   \label{lem:riczero}
%   Let $R_{abcd}$ be a curvature-like tensor with vanishing invariants.
%   If the Ricci condition \eqref{eq:rcond} \emph{is not} satisfied,
%   then there exists an aligned null direction $\ell^a$ along which the
%   boost order of $R_{abcd}$ is negative.
% \end{lemma}

\begin{lemma}
  \label{lem:symalign}
  Let $S_{abcd}$ be a symmetric, rank $4$ tensor with vanishing
  zeroth order invariants.  Then there exists an aligned null direction
  $\ell^a$; i.e. $S_{0000}=0$.
\end{lemma}

\begin{lemma}
  \label{lem:zrn}
  Let $R_{abcd}$ be a curvature-like tensor with vanishing zeroth order invariants.
  If $\supp{R}{2}_{\;\alpha\beta}=0$ and $R_{ab}=0$, then there exists
  an aligned null direction $\ell^a$ along which the boost order is
  negative.
\end{lemma}

\begin{lemma}
  \label{lem:zr2}
  Let $R_{abcd}$ be a curvature-like tensor with negative boost order.
  If $\supp{R}{2}_{\;\alpha\beta}=0$ and $R_{ab}=0$, then the principal type is N;
  i.e., \eqref{eq:ptypeN} holds.
\end{lemma}

\begin{proofof}{Theorem \ref{th:VSIcform}}
  If $R_{ab}\neq 0$, then the Theorem follows by Lemma \ref{lem:nzr}.
  Henceforth, we suppose that $R_{ab}=0$. If
  $\supp{R}{2}_{\;\alpha\beta}=0$, then the theorem follows by Lemma
  \ref{lem:zrn} and Lemma \ref{lem:zr2}.  Henceforth, we suppose that
  $\supp{R}{2}_{\;\alpha\beta}\neq0$.

  By Proposition \ref{prop:rank2VSI} of the Appendix,
  $R_{\alpha\beta}$ is nilpotent. Hence, for some sufficiently large
  $k$ we have
  $$P_{\alpha\beta} = \supp{R}{k}_{\;\alpha\beta}\neq 0,\quad
  \supp{P}{2}_{\;\alpha\beta}=0.$$
  We now consider
  two cases, depending on whether the  covariant
  $P_{ab} = P_{acb}{}^c$
  is non-zero or whether it vanishes.
    In the first case $P_{ab}\neq 0$, and we proceed as we did
  in the proof to Lemma \ref{lem:nzr} by constructing a rank 2
  covariant
  $$Q_{ab} = \ell_a \ell_b,$$
  and then showing that the boost order is
  negative.  If $P_{ab}=0$, then, we apply Lemma \ref{lem:zrn} and
  Lemma \ref{lem:zr2} to $P_{abcd}$ to find an aligned $\ell^a$ such
  that
  $$P_{abcd} = 4 P_{1i1j} \lmlm{i}{j}.$$

  Consider the following rank-4 covariant
  $$Q_{abcd} =  P_{aebf}
  P_c{}^e{}_d{}^f=\lambda \ell_a \ell_b \ell_c \ell_d,\quad \lambda =
  \sum_{ij} \lp P_{1i1j}\rp^2 >0.$$
  Proceeding as in the proof of Lemma \ref{lem:nzr}, we can show that
  $$R_{0i01}=R_{0ijk}=0.$$
  Since components of negative boost weight cannot
  contribute to an invariant, Lemma \ref{lem:pwzero} implies that the
  weight-zero components of $R_{abcd}$ vanish.  The theorem is thus proven.
\end{proofof}

\section{Necessity Proof}
\label{sec-nec}

In this Section we prove the necessity part of Theorem
\ref{main-theorem} assuming that the Riemann tensor has negative
boost weight; i.e., the Weyl tensor is of type III, N or O and
the Ricci tensor is of type N or O as was shown in Theorem
\ref{th:VSIcform}.

We will explicitly express two differential Weyl invariants for
type N and III vacuum spacetimes. These differential invariants
were originally used for similar proofs in 4D
\cite{BiPr,VSI,Pravda} and, as in 4D, they are zero only if both
the expansion and twist vanish. Though the resulting expressions
are simple, the calculations are quite extensive even with the use
of {\sc Maple}.

We cannot repeat similar calculations
for non-vacuum spacetimes since, at present, the consequences of the Bianchi identities for non-vacuum
spacetimes have not been fully studied and thus we cannot employ possible non-vacuum analogs of (\ref{PsitypeN})--(\ref{vyslB}).
For non-vacuum spacetimes we thus prove existence of non-vanishing differential Ricci invariants if $\bl$ is not geodesic
or if it is expanding or twisting. The explicit form of these invariants may be reconstructed
from the proof if necessary.

\subsection{Type N vacuum spacetimes}

Curvature invariants of the zeroth and first order (i.e.,
invariants containing the Riemann tensor and its first covariant
derivative) vanish. Curvature invariants of the second order lead
to invariants of matrices $\Psi_{ij}$ (\ref{WeylIIIN}) and $\mL$
(\ref{Lij}). Let us explicitly calculate the second order
invariant \BE I=C^{a b c d  ; r s} C_{a m c  n ; r s} C^{t m u n ;
v  w} C_{t b u d  ; v w}
 \EE
given in \cite{BiPr}. This expression can be rewritten as a
polynomial invariant constructed from the components of the
matrices $\mPsi$ and $\mL$ which contain more than thousand of
terms. A typical term is, for example,  \BDM 2^7 \Psi_{i j}
\Psi_{i l} \Psi_{m r} \Psi_{k t} L_{j k} L_{l m} L_{n p} L_{n q}
L_{r p} L_{r q} L_{r s} L_{t s}, \EDM which, due to Lemma 1 in
\cite{Bianchi}, can be simplified to $2^7 p^2 (S^2+A^2)^4$. We
note that it is efficient to use part (f) of  this lemma  as often
as possible, and only afterwards to decompose $\mL$ into $\mS$ and
$\mA$ and use the remaining equations in Lemma 1. Extensive
algebraical calculations in {\sc Maple} lead to \BE I=3^2 2^{10}
p^2 (S^2+A^2)^4. \label{invar} \EE This invariant clearly vanishes
only if both quantities $S$ and $A$ are equal to zero, and we have
thus completed the necessity part of the proof for type N vacuum
spacetimes.

%Any second order invariant can be expressed as a polynomial
%in $p (\theta^2+\omega^2)^2$.

Let us  check this formula in 4D. The relations between the frame
vectors $\bm^{2}$ and $\bm^{3}$ and the standard null tetrad
vectors $\bm$ and ${\mbox{\boldmath{$\bar m$}}}$  are \BE m^a =
\frac{1}{\sqrt{2}}(\msup{2}{a} - i\, \msup{3}{a}) , \quad {\bar
{m}}^a = \frac{1}{\sqrt{2}}(\msup{2}{a} + i\, \msup{3}{a}). \EE In
4D, $\mPsi$  has two independent real components, $\Psi_{22}$ and
$\Psi_{23}$,
 which are related to  the complex $\psi_4 = C_{a bcd} n^a {\bar m}^b
 n^c {\bar m}^d$ by
\BE \psi_4 = 2 (\Psi_{22}+ i \Psi_{23}) , \quad {\bar \psi_4} = 2
(\Psi_{22}- i \Psi_{23}) , \quad \psi_4 {\bar \psi_4} = 4
({\Psi_{22}}^2 + {\Psi_{23}}^2) . \EE Now we can recover the
formula for the invariant $I$ in 4D \cite{BiPr}: \BE I_{4D} = 3^2
2^8 (\theta^2+\omega^2)^4 \psi_4^2 \bar \psi_4^2 , \label{invar4D}
\EE by using $S=\theta$, $A=\omega$, $p=\frac{1}{2} \psi_4 {\bar
\psi_4}$.

%Note that
%\begin{center}
%$\mPsi =$ $\left( \begin {array}{cc} \psi_{11}& \psi_{12}\\
%\noalign{\medskip}\psi_{12}&-\psi_{11}\end {array} \right)$, $\ \
%\mPsi^2=$$({\psi_{11}}^2+{\psi_{12}}^2)$$
%\left( \begin {array}{cc} 1&0\\\noalign{\medskip}0&1\end {array} \right) $ \\
%\end{center}
%and thus
%\BE
%p=\frac{1}{2} \psi_4 {\bar \psi_4} .
%\EE
%Substituting this formula into (\ref{invar}), we recover (\ref{invar4D}).

\subsection{Type III vacuum spacetimes}

The zeroth order curvature invariants again vanish;
 however, we can calculate the first order non-vanishing invariant
\BE
I_{{III}} = C^{abcd;e} C_{amcn;e} C^{lmrn;s} C_{lbrd;s}
\EE
given in  \cite{Pravda}. Due to Eqs. (\ref{vyslS})--(\ref{vyslB}), we can express
this invariant, after
extensive calculations in {\sc Maple},  as
\BE
I_{{III}} = 64 (S^2+A^2)^2 \left[ 9 \psi^2 +27 \psi
(\cO_{PP} + \cO_{PF}) + 28 (\cO_{PP} + \cO_{PF})^2 \right],
\label{Inv3}
\EE
where $\cO_{PP} = \cO_{P \alpha P} \cO_{P \alpha P}$, $\cO_{PF} = \cO_{P \alpha F} \cO_{P \alpha F}$.

We note that $\cO_{PP}$, $\cO_{PF}$, and $\psi$ are non-negative
and for type III spacetimes at least one of them is positive and
thus the VSI condition for type III vacuum spacetimes implies $S
=A = 0$. We also note that the solution
(\ref{vyslS})--(\ref{vyslB}) is expressed for simplicity in a
frame which is adapted to the twisting case with non-vanishing
$\Psi_i$. However, this solution may be also expressed in another
frame in which we obtain the non-twisting case by  simply putting
$\mA=0$ and the case with $\Psi_i=0$ by putting $\Psi_i=0$. Thus
we can  obtain the resulting expressions for the invariant
$I_{{III}}$ in these special cases by substituting $A=0$ or
$\psi=0$ in Eq. (\ref{Inv3}). We remark  that the  completeness of the
proof for the vacuum type III case relies on the solution
(\ref{vyslS})--(\ref{vyslB}) being a general solution. Although
there is good analytical evidence to support this for six
dimensions and there is some support in higher dimensions, we have
not rigorously proven this in all the degenerate cases (of measure
zero) for the twisting case in dimension six and higher.

\subsection{Ricci invariants}

In this section we  show that for non-vacuum PP-N and PP-O
spacetimes there exist non-vanishing first and second order (in
derivatives) Ricci invariants if $\bl$ is not geodesic or if it
admits expansion or twist. We start with proving several lemmas
and then we apply them to the PP-N and PP-O cases separately.

\begin{lemma}
\label{lemmapureb0} If there exists a null frame in which a tensor
${\mbox{\boldmath{$T$}}}$ of rank {\bf 2} is of pure
boost order zero, then ${\mbox{\boldmath{$T$}}}$ possesses a
non-vanishing invariant.
\end{lemma}%
\begin{proof}
We prove this lemma by contradiction by assuming that a
second rank tensor of pure boost order zero,
\BE
T_{ab} = T_{01} n_{a} \ell_{b}+T_{10} \ell_{a} n_{b}+ T_{ij} \msub{i}{a} \msub{j}{b},
\EE
has vanishing algebraical scalar invariants.
Consequently,
\BE
T^{(2)}_{ac} \equiv T_{ab} T_{c}^{\ b} = T_{01} T_{10} ( n_a \ell_c
+  \ell_{a}  n_c) +T_{is} T_{js}  \msub{i}{a} m^j_c
%(T_{01})^2 n_{a} l_{c} + (T_{10})^2 l_{a} n_{c} + T_{ij} T_{jk} \msub{i}{a} m^k_c
\EE
and
\BE
T^{(2)}_{ac} T^{(2) ac} = 2 (T_{01})^2 (T_{10})^2 + T^{(2)}_{ij} T^{(2)}_{ij},
\EE
where
\BE
T^{(2)}_{ij} = T_{si} T_{sj}.
\EE
Thus, if ${\mbox{\boldmath{$T$}}}$ has vanishing algebraical invariants then
\BEA
T^{(2)}_{ij} =0 \Rightarrow \ T_{ij} =0  \quad {\rm and} \quad
\ T_{01} T_{10}= 0.
\EEA
Now, considering that
\BE
T_{ab} T^{ba} = (T_{01})^2+(T_{10})^2+T_{ij} T_{ji}
\EE
we conclude that
\BE
T_{01} = T_{10}= 0
\EE
and thus a non-vanishing ${\mbox{\boldmath{$T$}}}$ cannot have vanishing algebraical invariants.
\end{proof}
\begin{lemma}
\label{lemmarank2} If there exists a null frame in which a tensor
${\mbox{\boldmath{$T$}}}$ of rank {\bf 2}  has  boost order
zero, then ${\mbox{\boldmath{$T$}}}$ possesses a non-vanishing
invariant.
\end{lemma}
\begin{proof}
Every tensor $ {\mbox{\boldmath{$T$}}}$ with boost order zero can be divided into two parts
\BE
{\mbox{\boldmath{$T$}}}= {\mbox{\boldmath{$T$}}}^{(0)} + {\mbox{\boldmath{$T$}}}^{(-)}
\EE
where ${\mbox{\boldmath{$T$}}}^{(0)}$ is of pure boost order zero and
${\mbox{\boldmath{$T$}}}^{(-)}$ has negative boost order.

Clearly ${\mbox{\boldmath{$T$}}}^{(-)}$  does not affect any
invariant of ${\mbox{\boldmath{$T$}}}$, and thus this Lemma is a
direct consequence of the Lemma \ref{lemmapureb0}.
\end{proof}

\begin{lemma}
\label{lemmarank3} If there exists a null frame in which a tensor
${\mbox{\boldmath{$T$}}}$ of rank {\bf 3}  has pure boost
order zero, then ${\mbox{\boldmath{$T$}}}$ possesses a
non-vanishing invariant.
\end{lemma}
\begin{proof}
We again prove this lemma by contradiction.
A general form of a third rank tensor of pure boost order zero is
\BEA
T_{abc} = T_{01i} n_{a} \ell_{b} m_{c}^{i} + T_{10i} \ell_{a} n_{b} m_{c}^{i} + T_{0i1} n_{a} m_{b}^{i} \ell_{c} +
T_{1i0} \ell_{a} m_{b}^{i} n_{c} \nonumber \\
+ T_{i01} m_{a}^{i} n_{b} \ell_{c} + T_{i10} m_{a}^{i} \ell_{b}
n_{c} + T_{ijk} m_{a}^{i} m_{b}^{j} m_{c}^{k}. \EEA Let us now
construct from ${\mbox{\boldmath{$T$}}}$ several second rank pure
boost order zero tensors. If ${\mbox{\boldmath{$T$}}}$ has
vanishing algebraical invariants, then all of these tensors have to
vanish according to Lemma \ref{lemmapureb0}. In fact, we do not
need to express these tensors fully,  we  just need some of their
components: \BEA
T_{abc} T_{d}^{\ cb}  \ell^{a} n^{d} = T_{01i} T_{1i0} + T_{0i1} T_{10i}, \label{rank31}\\
T_{abc} T^{a \ b}_{\ d} \ell^{c} n^{d} = T_{01i} T_{1i0} + T_{i10} T_{i10}, \\
T_{abc} T^{a \ b}_{\ d} n^{c} \ell^{d} = T_{0i1} T_{10i} + T_{i01} T_{i01}. \label{rank33}
\EEA
An appropriate linear combination of Eqs. (\ref{rank31})--(\ref{rank33})  leads to
\BE
T_{i10} T_{i10} + T_{i01} T_{i01} = 0
\EE
and thus
\BE
T_{i10} = T_{i01} = 0. \label{rank3van1}
\EE
Other components of pure boost order zero tensors of rank 2 are
\BEA
T_{abc} T^{b \ c}_{\ d} \ell^{a} n^{d} = T_{01i} T_{01i} + T_{0i1} T_{i10}, \label{rank34} \\
T_{abc} T^{b \ c}_{\ d} n^{a} \ell^{d} = T_{10i} T_{10i} + T_{1i0} T_{i01}, \label{rank35} \\
T_{abc} T_d^{\ ba} \ell^{c} n^{d} = T_{1i0} T_{1i0} + T_{i10} T_{10i}, \label{rank36} \\
T_{abc} T_d^{\ ba} n^{c} \ell^{d} = T_{0i1} T_{0i1} + T_{i01} T_{01i}. \label{rank37}
\EEA
From the vanishing of Eqs. (\ref{rank3van1})--(\ref{rank37}), it follows that
\BE
T_{01i} = T_{10i} = T_{1i0} = T_{0i1} =0. \label{rank3van2}
\EE
Now we need only one last invariant
\BE
T_{abc} T^{abc} = 2 T_{01i} T_{10i} + 2 T_{0i1} T_{1i0} + 2 T_{i01} T_{i10} + T_{ijk} T_{ijk}
\EE
which thanks to Eqs. (\ref{rank3van1}) and (\ref{rank3van2}) vanishes only if
\BE
T_{ijk} = 0.
\EE
Thus if ${\mbox{\boldmath{$T$}}}$ has vanishing algebraical invariants then
${\mbox{\boldmath{$T$}}}$ vanishes.
\end{proof}

\subsubsection{PP-N spacetimes}

For PP-N  spacetimes the Ricci tensor has the form \cite{cmpppz}
\BE
R_{ab} = K_i (\ell_a m^i_b + \msub{i}{a} \ell_b
) + A \ell_a \ell_b, \label{PPN}
\EE
with $K_i \not= 0$ for at least one value of $i$.

Now a tensor of rank 3 \BE T_{a b c } \equiv R_{a b; \sigma }
R^\sigma_{\ c} \label{T1} \EE has boost order zero, and thus
(Lemma \ref{lemmarank3}) for VSI spacetimes all of its components
with boost weight zero have to vanish. By expressing the component
\BE T_{ijk} =  K_{i} L_{j0}  K_k + L_{i0}  K_j K_k, \EE and
multiplying this equation by $L_{i0}$ and by contracting $k$ with
$j$, we obtain \BE (K_{i}  L_{i0})^2   + L_{i0} L_{i0}    K_j K_j
= 0 \EE which implies \BE L_{i0}    = 0 ;\label{Li20} \EE i.e., $\bl$
is geodesic, and thus also \BE \Mi_{00}=0.
\label{Mi220} \EE Assuming that Eqs. (\ref{Li20}) and (\ref{Mi220}) are
satisfied, we find that the first covariant derivative of the
Ricci tensor $R_{a b; c }$ has the boost order 0 and thus for VSI
spacetimes all of its components of boost weight 0 must vanish.
From \BE R_{ij;k} = L_{jk} K_{i} + L_{ik} K_{j} = 0, \label{rceLij}
\EE and contracting $i$ with $j$, it follows that \BE L_{ik} K_i =
0. \label{rceLijpom} \EE Multiplying Eq. (\ref{rceLij}) by $K_j$ and
using (\ref{rceLijpom}) leads to \BE K_j K_j L_{ik} = 0 \EE which
implies that \BE L_{ij}=0; \EE i.e., the expansion and twist matrices are
zero.

\subsubsection{PP-O  spacetimes}

For PP-O  spacetimes the Ricci tensor has the form \cite{cmpppz}
\BE R_{ab}=A \ell_a \ell_b. \EE The covariant derivative
$R_{ab;c}$, has boost order 0. The boost weight zero component \BE
R_{1i;0} = A L_{i0} \EE has to vanish from the VSI condition and
thus \BE L_{i0}=0 \EE and $\bl$ is geodesic. We choose the affine
parametrization with $L_{10}=0$. Now, the second rank tensor
$R_{ab;c}^{\ \ \ \ c}$ has boost order 0. Thus the component with
boost weight 0 \BE R_{ij;\sA}{}^{\!\sA} = 2 A L_{jk} L_{ik} \EE
has to vanish from the VSI condition and consequently \BE
L_{ij}=0, \EE i.e., the expansion and twist matrices are zero.

\section{\bf Discussion}

We have proven that in N-dimensional Lorentzian spacetimes the
boost order of the Riemann tensor is negative along some aligned
non-expanding, non-twisting,  geodesic null direction $\ell^a$ if
and only if all scalar curvature invariants vanish (generalizing a
previous theorem in 4D \cite{VSI}). We emphasize that even though
our main focus has been the Lorentzian geometry case, some results
from the more general theory ($N$-dimensional, real vector space
equipped with an inner product $g_{ab}$ with no assumption about
the signature) have been developed.

In the Lorentzian case, we have provided strong evidence for the
following conjecture ({\em the algebraic VSI conjecture}): any tensor
with vanishing algebraic scalar invariants must necessarily have
negative boost order along some aligned null direction.  We have
given a proof of this conjecture for arbitrary dimensions and for
the following tensor types: bivectors in Proposition
\ref{prop:vsibv}, symmetric rank 2 tensors in Corollary
\ref{cor:pnform}, and for the general class of curvature-like
tensors in Theorem~\ref{th:VSIcform}.  Lemmas \ref{lemmarank2} and
\ref{lemmarank3} provide additional evidence for this conjecture.

We note that all of the VSI spacetimes have a shear-free,
non-expanding, non-twisting geodesic null congruence $\bl=\partial_v$, and
hence belong to the "generalized Kundt" class \cite{cmpppz}.

There is a number of potentially important physical applications
of VSI~ spacetimes. For example, it is known that a wide range of
VSI~spacetimes (in addition to the pp-wave spacetimes
\cite{amati,HS}) are exact solutions in string theory (to all
perturbative orders in the~string tension) \cite{coley}. Recently,
type-IIB superstrings in pp-wave backgrounds with an ~RR five-form
field were also shown to be exactly solvable \cite{matsaev}.
Indeed, many authors \cite{amati,tseytlin} have investigated
string theory in pp-wave backgrounds in order to search for
connections between quantum gravity and gauge theory dynamics.

In the context of string theory, it is of considerable interest to
study Lorentzian spacetimes in higher dimensions. In particular,
higher dimensional generalizations of pp-wave  backgrounds have
been considered \cite{tseytlin,rutone}, including string models
corresponding not only to the NS-NS but also to certain R-R
backgrounds \cite{RT,russot}, and pp-waves in eleven-
and ten-dimensional supergravity theory \cite{GuevenAD}. In
addition,  a number of classical solutions of branes \cite{brane}
in higher dimensional pp-wave backgrounds have been studied in
order to better understand the non-perturbative dynamics of string
theories. In particular, a class of pp-wave string spacetimes
supported by non-constant NS-NS $H_3$  or R-R $F_p$ form fields
were shown to be exact type II superstring solutions to all orders
in the string tension \cite{mam,russot}. In this class of
10-dimensional superstring theory models the pp-wave metric, the
NS-NS 2-form potential and the 3-form  $H_3$ background, which
depends on arbitrary  harmonic functions $b_m(x)$ ($\del^2 b_m=0,
\ m=1,2,...$) of the transverse  coordinates $x_i$, are given by
\cite{russot} \BE ds^2=dudv + K(x)du^2+dx_i^2+dy_m^2\ , \ \ \ \ \
\ \ i=1,...,d\ , \ \ \ m=d+1,..,8\ , \EE \BE B_2= b_m (x)\
du\wedge dy_m \ , \  \ \ \ \ \ H_3 = \del_i  b_m(x)\ dx_i\wedge
du\wedge dy_m\, \EE  where the only non-zero component of the
generalized curvature is \BE \hat R_{uiuj} = - \ha \del_i \del_j K
- \ha  \del_i b_m \del_j b_m \ . \EE
These solutions are consequently of PP-type O and of principal (algebraic
Weyl) type N.

{\em Acknowledgements}.
We would like to thank Jose Senovilla for helpful comments.
 VP and AP would like to thank Dalhousie University for  hospitality while
this work was  carried out. AC and RM were supported
 by research grants from NSERC, VP was supported by research grant GACR-202/03/P017 and AP by
grant KJB1019403.

\appendix
\section{Appendix: Indefinite Signature Inner Products}

Let $0<p<q$ be integers, and let $\reals^{p,q}$  denote
$\reals^{p+q}$ equipped with a signature $(p,q)$ inner product
$$\sum_{\lambda=1}^p X^{\lambda} Y^{\lambda} -
\sum_{\lambda=p+1}^{p+q} X^{\lambda}Y^{\lambda},\quad
X^\lambda,Y^\lambda\in\reals^{p+q}.$$
% We use
% $\lambda,\mu=1,\ldots,p+q$ to denote component indices, and use
% $\ovEsub{\lambda}{i},\, i=1,\ldots,p+q$ to denote the standard basis
% of $\reals^{p+q}$.
Let $\SO_{p,q}$ be the corresponding group of
$(p,q)$-orthogonal transformations.  We consider a collection
$\Xsup{\lambda}{i}$ of mutually orthogonal null vectors.  In other
words, for all values of collection indices $i,j$ we have
\begin{equation}
  \label{eq:nvortho}
  \sum_{\lambda=1}^p \Xsup{\lambda}{i}\Xsup{\lambda}{j} -
  \sum_{\lambda=p+1}^{p+q} \Xsup{\lambda}{i}\Xsup{\lambda}{j}=0.
\end{equation}
\begin{proposition}
  \label{prop:indefli}
  A collection of mutually orthogonal
  null vectors, $\Xsup{\lambda}{i}\in \reals^{p,q}$,
  has at most $p$ linear independent elements.
\end{proposition}
\begin{proposition}
  Let $\Xsup{\lambda}{i}\in \reals^{p,q}$ be a collection of $p$,
  or fewer, linear independent, mutually orthogonal null vectors.
  Then, there exists a $(p,q)$ isometry $T^{\lambda}_{\;\;\mu}\in
  \SO_{p,q}$ such that the transformed collection of mutually
  orthogonal vectors
  $$\Ysup{\lambda}{i} =  T^\lambda_{\;\;\mu}\, \Xsup{\mu}{i}$$
  has the form $\Ysup{i}{i} = \Ysup{i+p}{i} = 1$, with all other
  components $0$.
\end{proposition}
\begin{corollary}
  \label{cor:indefcoll}
  Let $\Xsup{\lambda}{i}\in \reals^{p,q}$ be an arbitrary
  collection of mutually orthogonal null vectors.
  Then, there exists a $(p,q)$ isometry $T^{\lambda}_{\;\;\mu}\in
  \SO_{p,q}$ such that the transformed collection of mutually
  orthogonal vectors
  $$\Ysup{\lambda}{i} =  T^\lambda_{\;\;\mu}\, \Xsup{\mu}{i}$$
  satisfies
  $$\Ysup{\lambda}{i} = \Ysup{\lambda+p}{i},\;
  \lambda=1,\ldots,p,\quad\mbox{and}\quad \Ysup{\lambda}{i}=0,\;
  \lambda>2p.$$
\end{corollary}

\section{Appendix: the Petrov Normal Form}
\label{ap:pnf}

Even though our focus is Lorentzian geometry, some
signature-independent results need to be developed.  Let $g_{\delta
  \epsilon}$ be an $N$-dimensional, non-degenerate inner product; we make
no assumptions about signature\footnote{As a notational reminder we
  will use $\delta, \epsilon,\gamma$ to index tensors in the more general
  setting, and reserve $a,b,c$ as indices of tensors in a Lorentzian
  setting.}. Let $\tT=T_{\delta\epsilon}$ be a general, rank 2 tensor.
For each $k=1,2,\ldots, N-1$ let
$$\sigma_k = \supp{T}{k}{}_\epsilon{}^\epsilon =
T_{\epsilon_1}^{\;\epsilon_2}\, T_{\epsilon_2}^{\;\epsilon_3}\, \cdots \,
T_{\epsilon_k}^{\;\epsilon_1}$$
denote the $k\supth$ power invariant of
$\tT$.  The following is well known \cite{fulhar}:
\begin{theorem}
  Every scalar invariant of $\tT$ has a unique representation as a
 polynomial of the power invariants $\sigma_0,\ldots,\sigma_{N-1}$.
\end{theorem}
\begin{corollary}
  \label{prop:rank2VSI}
  A rank 2 tensor $\tT$ has vanishing zeroth order invariants (i.e., is $VSI_0$) if and
  only if it is nilpotent; i.e., $\supp{\tT}{k}=0$ for some $k\geq0$.
\end{corollary}

Next, we consider a symmetric, rank 2 tensor
$Q_{\delta\epsilon}=Q_{\epsilon\delta}$.  The normal forms
described below are a specialization of the normal forms described
by Petrov  \cite{petrov}, and the proof of the following result
can be found therein.
\begin{theorem}
  \label{th:petrovnform}
  If a symmetric tensor $Q_{\delta\epsilon}$ belongs to the $VSI_0$ class, then
  there exists a basis,
  $$\ovdK{\lambda,\nu}{\epsilon},\quad \lambda=1,\ldots,r;\quad
  \nu=1,\ldots,d_\lambda\;,$$
  and a sequence of block signatures
  $\sigma_\lambda = \pm 1,$
  such that
  \begin{eqnarray}
    \label{eq:pnf1}
    g_{\delta\epsilon} &=& \sum_{\lambda,\nu}
    \sigma_\lambda\ovdK{\lambda,\nu}{\!(\delta}\,
    \ovdK{\lambda,\rho}{\!\epsilon)}  ,    \\
    \label{eq:pnf2}
    Q_{\delta\epsilon} &=& \sum_{{\lambda,\nu\atop d_\lambda>1}}
    \sigma_\lambda\! \ovdK{\lambda,\nu+1}{\!\!\!(\delta}\,
    \ovdK{\lambda,\rho}{\!\epsilon)} ,
  \end{eqnarray}
  where $\lambda=1,\ldots,r$, where $\nu=1,\ldots, d_\lambda$, and
 we are letting $\rho=d_\lambda +1-\nu$.
\end{theorem}
\begin{corollary}
  \label{cor:vsiq2}
  Let $Q_{\delta\epsilon}$ be a symmetric, rank 2 tensor. If
  $\suppsub{Q}{2}{\delta\epsilon}=0$, then
  $$Q_{\delta\epsilon} = \sum_{\lambda=1}^{p}
  \ovdK{\lambda}{\delta}\ovdK{\lambda}{\epsilon} - \sum_{\lambda=p+1}^{p+q}
  \ovdK{\lambda}{\delta}\ovdK{\lambda}{\epsilon},
  $$
  where the $\ovdK{\lambda}{\epsilon},\; \lambda=1\,\ldots,p+q,$ are
  mutually orthogonal
  null vectors; i.e.,
  $ \ovdK{p}{\epsilon}\ovuK{q}{\epsilon}=0.$
\end{corollary}
\begin{proof}
   Conditions \eqref{eq:pnf1} and \eqref{eq:pnf2} are equivalent to
   the statement that the linear
   transformation $Q_\delta{}^\epsilon$ has the action
   \begin{equation}
     \label{eq:eblock-action}
     \ovdK{\lambda,1}{\epsilon} \to \ovdK{\lambda,2}{\epsilon}\to \ldots
     \to \ovdK{\lambda,d_\lambda}{\epsilon} \to 0.
   \end{equation}
   Since $\suppsub{Q}{2}{\delta\epsilon}=0$, we have that $d_\lambda=1$
   or $2$ for all $\lambda$.  We then rearrange  our basis so that
   $\sigma_\lambda=1, d_\lambda=2$ for $\lambda=1,\ldots, p$, so that
   $\sigma_\lambda=-1, d_\lambda=2$ for $\lambda=p+1,\ldots, p+q$, and
   so that $d_\lambda=1$ for $\lambda>p+q$.  We obtain the desired
   form by setting
   $$\ovdK{\lambda}{\epsilon}= \ovdK{\lambda,2}{\epsilon},\quad
   \lambda=1,\ldots, p+q.$$
\end{proof}
% Note that the first condition is equivalent to the statement that the
% dual basis is given by
% $$\ovu{\tK}{p}{q} = \sigma_p \ovdtK{p}{d_p+1-q}.$$

Conditions \eqref{eq:pnf1} and \eqref{eq:pnf2} are equivalent to the
assertion that $g_{\delta\epsilon}$ and $Q_{\delta\epsilon}$ can be
simultaneously put into block
diagonal form:
$$
g_{\delta\epsilon} =
\pmatrix{
  \ovG{1} & & \cr
  & \ddots \cr
  & & \ovG{r}}
\qquad
Q_{\delta\epsilon} =
\pmatrix{
  \ovQ{1} &&\cr
  & \ddots\cr
  && \ovQ{r}},
$$
such that the blocks have the form
$$
\ovG{\lambda}  =
\pmatrix{
  && \sigma_\lambda \cr
  & \addots \cr
  \sigma_\lambda},\qquad
\ovQ{\lambda} =
\pmatrix{
 && &\sigma_\lambda & 0 \cr
 &&\sigma_\lambda & 0 \cr
  & \addots & \addots \cr
  \sigma_\lambda & 0 \cr
  0}.
$$
Thus, there are four kinds of blocks, depending on the block
signature $\sigma_p$ and on the block parity --- a block being
called even or odd according to whether $d_\lambda$ is even or
odd.  The basis vectors of an even block are pairs of conjugate
null vectors. The same is true for odd blocks, save that the
middle vector is either a unit space-like, or a time-like vector
depending on whether the signature $\sigma_\lambda$ is $+1$ or
$-1$, respectively.  The following table summarizes the four
possibilities, and the corresponding signatures of the blocks:
\begin{center}
  \renewcommand{\arraystretch}{1.2}
  \begin{tabular}{cccc}
    $\sigma_\lambda$ & parity & sig. $\ovG{\lambda}$ & sig. $\ovQ{\lambda}$ \\
    \hline
    $+1$ & even & $(\frac12 d_\lambda,\frac{1}{2} d_\lambda)$ &
    $(\frac{1}{2}d_\lambda,
    \frac{1}{2}d_\lambda-1)$ \\
    $-1$ & even & $(\frac{1}{2}d_\lambda,\frac{1}{2}d_\lambda)$ &
    $(\frac{1}{2}d_\lambda-1, \frac{1}{2}d_\lambda)$ \\
    $+1$ & odd & $(\frac{1}{2}(d_\lambda+1),\frac{1}{2}(d_\lambda-1))$ &
    $(\frac{1}{2}(d_\lambda-1), \frac{1}{2}(d_\lambda-1))$ \\
    $-1$ & odd & $(\frac{1}{2}(d_\lambda-1),\frac{1}{2}(d_\lambda+1))$ &
    $(\frac{1}{2}(d_\lambda-1), \frac{1}{2}(d_\lambda-1))$
  \end{tabular}
\end{center}
Summing the signatures of all the blocks we arrive at the following result.
\begin{proposition}
  \label{prop:vsig}
  Suppose that a symmetric $Q_{\delta\epsilon}$ has vanishing zeroth
  order invariants (i.e., is $VSI_0$).  Then, the signature of the inner product
  $g_{\delta\epsilon}$ is given by
  \begin{equation}
    \label{eq:Asig}
    \left(\frac{N+\sigma_o}{2}\,,\,\frac{N-\sigma_o}{2}\right),
    \quad\mbox{where}\quad
    \sigma_o=\sum_{d_\lambda\mathrm{\ odd}}  \sigma_\lambda;
  \end{equation}
  and the signature of $Q_{\delta\epsilon}$ by
  \begin{equation}
    \label{eq:Bsig}
    \left(\frac{N-\nu+\sigma_e}{2}\,,\,
    \frac{N-\nu-\sigma_e}{2}\right),
    \quad\mbox{where}\quad \sigma_e=\sum_{d_\lambda\mathrm{\ even}}
    \sigma_\lambda.
  \end{equation}
\end{proposition}

As a particular case of the above proposition, suppose that $g_{ab}$
has Lorentz signature, $(N-1,1)$.  In this case, Eq.
\eqref{eq:Asig} is a very strong constraint on the size and number of
odd and even blocks.  Indeed, there can be at most one block of size 2
and signature $(1,1)$, or one block of size 3 and signature $(2,1)$.
The possibilities are summarized below.
\begin{corollary}
  \label{cor:pnform}
  Suppose that $g_{ab}$ has Lorentz-signature. Then, there exists a
  null-frame $\ell^a,n^a,m_i{}^{a}$ relative to which a
  $VSI_0$ $Q_{ab}$ takes
  on exactly one of the following normal forms:
  \begin{eqnarray}
    \label{eq:qnf1}
    Q_{ab}&=&0;\\
    \label{eq:qnf2}
    Q_{ab}&=&\pm \ell_a \ell_b;\\
    \label{eq:qnf3}
    Q_{ab}&=&\ell_{(a} m^2{}_{b)}.
  \end{eqnarray}
\end{corollary}

\section{Appendix: Bivectors}
\label{ap:bivectors}

Henceforth, we assume that $g_{ab}$ has Lorentz signature.
A bivector is a rank $2$ skew-symmetric tensor.  The vector inner
product $g_{ab}$ naturally induces a bivector inner product
$$g_{\alpha\beta}=\tfrac{1}{2}(g_{ac}\,g_{bd}-g_{ad}\,g_{bc}),\quad
\alpha=(a,b),\; \beta=(c,d).$$
We use $\alpha=(a,b),\; a<b$ to denote
a bivector index, and henceforth use $\alpha,\beta,\gamma,\ldots$ to
denote bivector indices.
A bivector index can take on $N(N-1)/2$
possible values; this is the dimension of the vector space of all
bivectors.
The inner product of two
bivectors $J_\alpha=J_{ab}$, $K_\alpha=K_{ab}$ can also be
characterized as the total contraction
$$ g_{\alpha\beta}\, J^\alpha K^\beta =J_\alpha K^\alpha = J_{ab} K^{ab}.$$

A  bivector $K_\alpha$ will be called null if  $K_\alpha K^\alpha=0$.
Every null-frame $\ell^a,\ n^a,\ m_i{}^a$ induces
a basis of bivectors consisting of
the $N-2$ pairs of conjugate, null bivectors
$\ell_{[a}m^i{}_{b]}, \; n_{[a}m^i{}_{b]},$
the negative-norm bivector
$\ell_{[a} n_{b]},$
and $\frac{1}{2}(N-2)(N-3)$ positive-norm bivectors
$m^i_{[a} m^j{}_{b]}$.
It follows that the bivector inner-product
$g_{\alpha\beta}$ has signature
$(\tfrac{1}{2}(N-1)(N-2),N-1)$.

%%%%%%%%%%%%%%%%%%%%%%%%%%%%%%

It can be shown \cite{MilSPT} that 
for even $N$ 
a bivector $K_\alpha$ always admits at least one
aligned, real, null direction, while for odd $N$ it
is possible that there is no real
aligned, null direction.
The boost weights of the components of $K_{ab}$ are given by 
\[  K_{ab} = \overbrace{2K_{0i}\, n_{[a} m^i{}_{b]}}^1 +
  \overbrace{2K_{01}\, n_{[a}\ell_{b]}+ K_{ij}\, m^i{}_{[a}
    m^j{}_{b]}}^0+
  \overbrace{2K_{1i}\, \ell_{[a} m^i{}_{b]}}^{-1}.  \]
For an aligned (or singly aligned) bivector 
we can set $K_{0i}=0$  (but $K_{1i}$ is not zero) and for a bi-aligned bivector 
we can set $K_{0i}=K_{1i}=0$. 
We can classify bivectors into alignement types
(using notation consistent with \cite{Algclass,Bianchi}).
We will say that a bivector is 
of type $G$ if $K_{0i}$ cannot be made to vanish,
of type $I$ if $K_{0i}=0$ (but $K_{01}$ does not vanish), and
of algebraically special type $II$ if $K_{0i}=K_{01}=K_{ij}=0$.
For type $II$, the  bivector is aligned and  
$K_{1i}$ cannot be made to vanish; i.e., there is no bi-aligned subclass.
For type $I$, we will say that the bivector is  bi-aligned 
if $K_{0i}=K_{1i}=0$, and we shall refer to this case as  
type $I_i$ (this case is akin to type $II_{ii}$ in the classification of the Weyl tensor
\cite{newcqg} and could perhaps also be referred to as type $D$). We also note that in 4D,
types $I_i$ and $II$
are referred to as types  $I$ and $N$, respectively \cite{stewart}.

%%%%%%%%%%%%%%%%%%%%%%%%%%%

A collection of bivectors $\ovdK{\lambda}{\alpha}$ is null and
mutually orthogonal if and only if
\begin{equation}
  \label{eq:mutortho}
  2\ovdK{\lambda}{0i} \ovduK{\mu}{1}{i} + 2\ovdK{\lambda}{1i}
  \ovduK{\mu}{0}{i}+
  \ovdK{\lambda}{ij} \ovuK{\mu}{ij}-2\ovdK{\lambda}{01} \ovdK{\mu}{01}= 0
\end{equation}
for all $\lambda, \mu$.
\begin{proposition}
  \label{prop:nullbv}
  Let $\ovdK{\lambda}{\alpha},\; \lambda=1,\ldots, r$ be a collection
  of null, mutually orthogonal, linearly independent bivectors.  Then,
  $r\leq N-1$.
\end{proposition}
\begin{proof}
  This follows directly from Proposition \ref{prop:indefli}.
\end{proof}
\begin{proposition}
  \label{prop:aligned-ortho}
  Let $\ovdK{\lambda}{\alpha}$ be a collection
  of null, mutually orthogonal bivectors with a common
  alignment, i.e., $\ovdK{\lambda}{0i}=0$.  Then,
  there exists a null, bi-aligned bivector $M_\alpha$ and
  scalars $C_\lambda$ such that
  $\ovdK{\lambda}{ij} = C_\lambda  M_{ij}$, and $\ovdK{\lambda}{01} =
  C_\lambda M_{01}$.
\end{proposition}
\begin{proof}
  Consider the sequence of type $I_i$ bivectors defined by
  $$\ovdM{\lambda}{0i}=0,\quad
  \ovdM{\lambda}{01} = \ovdK{\lambda}{01},\quad
  \ovdM{\lambda}{ij} = \ovdK{\lambda}{ij},\quad
  \ovdM{\lambda}{1i}=0.$$
  For a fixed $\ell^a$, the vector space of type $I_i$ 
  bivectors has signature $(1,(N-2)(N-3)/2)$.
  Hence, by Proposition \ref{prop:indefli}, the
  $\ovdM{\lambda}{\alpha}$ must be multiples of one another.
\end{proof}

\begin{proposition}
  \label{prop:vsibv}
  A bivector $K_\alpha$ has vanishing zeroth order scalar invariants
($VSI_0$) if and only if
  it is of alignment type $II$.
\end{proposition}
\begin{proof}
  Since an invariant has boost-weight zero, a $K_{ab}$ with negative
  boost order must have vanishing scalar invariants.  Let us now prove
  the converse;  i.e., we assume that $K_{ab}$ has vanishing scalar
  invariants and prove that necessarily the boost order is negative.

  Let us consider the symmetric rank 2 covariant
  $$R_{ab}=K_{ac} K_b^{\;c}.$$
  By Corollary \ref{cor:pnform}, we can choose an aligned $\ell^a$ so
  that
  $$R_{00}=R_{0i}=R_{01}=R_{ij}=0.$$
  Since
  $$ R_{00} = \sum_i (K_{0i})^2,$$
  we have that $K_{0i}=0$.  From
  $$ R_{ii} = -2K_{0i} K_{1i} +\sum_k (K_{ik})^2,\quad
    R_{01} = K_{01}^2-\sum_i K_{0i} K_{1i}.
  $$
  we infer that $K_{ij}=K_{01}=0$, as was to be shown.
\end{proof}

We note that these results may be of importance in the study
of  higher dimensional spacetimes with Maxwell-like fields \cite{GSW}.

\section{Appendix: The Proofs of the Lemmas}
\label{ap:lemmasA}

\begin{proofof}{Lemma \ref{lem:pwzero}}
  Let $R_{abcd}$ be a curvature-like tensor with terms of zero boost
  weight only.  We argue by contradiction, and suppose that $R_{abcd}$
  has vanishing scalar invariants.  By proposition \ref{prop:rank2VSI},
  $$\suppsub{R}{k}{abcd}=0$$
  for a sufficiently large $k$.
  Consequently, $\suppsub{R}{j}{abcd}$ has vanishing zeroth order
  invariants for all $j<k$.  Since $\suppsub{R}{j}{abcd}$ is of pure
  boost weight zero we may, without loss of generality, suppose that
  $\suppsub{R}{2}{abcd} = 0$.

  We decompose the curvature-like tensor as follows:
  $$R_{abcd}=A_{abcd}+B_{abcd},$$
  where
  \begin{eqnarray*}
    A_{abcd} &=& 8 R_{0101} \nlnl+  4 R_{01ij} \nlmm{i}{j} +
    R_{ijkl} \mmmm{i}{j}{k}{l};\\
    B_{abcd} &=& 8 R_{0i1j}\nmlm{i}{j}.
  \end{eqnarray*}
  Evidently,
  $$A_{\alpha\beta} B^\beta{}_\gamma = B_{\alpha\beta}
  A^\beta{}_\gamma = 0,$$
  and hence,
  \begin{equation}
    \label{eq:a2b2zero}
    \suppsub{A}{2}{abcd} =
    \suppsub{B}{2}{abcd}=0.
  \end{equation}
Now, there are two cases to consider; either $A_{\alpha\beta}$
  vanishes, or it
  does not.  If it does vanish, then
  $$R_{abcd} R^{adcb} = 4 \sum_{ij} \lp R_{0i1j} \rp^2$$
  is  a
  non-vanishing invariant.  Thus, without loss of generality,
  $A_{\alpha\beta}\neq 0$.

  Note that $A_{\alpha\beta}$ is a quadratic combination of type $I_i$  
  bivectors.  In Appendix \ref{ap:bivectors}, we showed that the
  vector space of type $I_i$  bivectors has Lorentz signature.  Hence,
  by Eqs. \eqref{eq:a2b2zero} and Corollary \ref{cor:vsiq2},
  $$A_{\alpha\beta} = \pm K_\alpha K_\beta,$$
  where $K_\alpha$ is a null, type $I_i$  bivector; i.e.,
  $$-2 \lp K_{01}\rp^2 + K_{ij} K^{ij} = 0.$$
  Consequently, $K_{01}\neq 0$, and hence,
  $$R_{0101}=A_{0101} = \pm \lp K_{01}\rp^2 \neq 0.$$

  The matrix
  $X_{ij} = B_{0i1j}$ is nilpotent by \eqref{eq:a2b2zero},
  and hence,
  $X_i^{\;i}=0$.  Let
  $R_{ab}=R_{acb}{}^c$ be the  Ricci covariant.
  We have
  $$R_{01} = -R_{0101} + R_{0i1}{}^i = -A_{0101} +X_i{}^i \neq 0.$$
  Hence
  $$R_{ab} R^{ab} = \lp R_{01}\rp^2 + R_{ij}R^{ij} $$
  is a
  non-vanishing invariant. We have established a contradiction and hence
  proved the lemma.
\end{proofof}

\begin{proofof}{Lemma \ref{lem:nzr}}
  Corollary \ref{cor:pnform} gives normal forms for $R_{ab}$ with
  vanishing invariants.  We choose an aligned $\ell^a$ so that
  \begin{equation}
    \label{eq:nzrbw-1}
    R_{00}=R_{0i}=R_{01}=R_{ij}=0.
  \end{equation}
  If $R_{ab} = 0$, then all contractions vanish. Assuming that
 $R_{ab}\neq 0$, we can construct
  a non-vanishing covariant of the form 
  $$Q_{abcd} = \ell_a \ell_b \ell_c \ell_d.$$
  To obtain this covariant
  we use $R_{ab}R_{cd} $ or $R_{ae} R^e_{\;b} R_{cf} R^f_{\; d}$,
  depending on whether $R_{ab}$ has the form \eqref{eq:qnf2} or the form
  \eqref{eq:qnf3}, respectively.  The vanishing of the
  invariant
  $$Q^{abcd} R_{aebf}R_c{}^e{}_d{}^f =  \sum_{ij} \lp R_{0i0j}\rp^2,$$
  implies that
  \begin{equation}
    \label{eq:nzrbw1}
    R_{0i0j}=0.
  \end{equation}

  We set $T_{\alpha\beta}=\suppsub{R}{2}{\alpha\beta}$ and note that  the
  vanishing of the invariant
  $$Q^{abcd} T_{aebf}T_c{}^e{}_d{}^f = \sum_{ij} \lp T_{0i0j}\rp^2$$
  implies that
  \begin{equation}
    \label{eq:nzrtbw1}
    T_{0i0j} = R_{0iab} R_{0j}{}^{ab} = 0.
  \end{equation}

  We define the following sequence of bivectors
  $$\ovdK{i}{ab} = R_{0iab};$$
  these are aligned because of Eq. \eqref{eq:nzrbw1}.
  Also, by Eq. \eqref{eq:nzrtbw1}, we have that
  $$T_{0i0j} = \ovdK{i}{\alpha} \ovuK{j}{\alpha}=0.$$
  By Proposition \ref{prop:aligned-ortho}, there exist $M_{jk},
  M_{01}$ and $C_i$ such that
  \begin{equation}
    \label{eq:mortho}
    M_{jk} M^{jk} - 2 M_{01}{}^2 =0,
  \end{equation}
  and such that
  \begin{equation}
    \label{eq:zr2bw-1}
    R_{0ijk}=\ovdK{i}{jk} = C_i M_{jk},\quad R_{0i01}=\ovdK{i}{01} =
    C_i M_{01}.
  \end{equation}
  By  Eq. \eqref{eq:nzrbw-1},
  $$R_{0j} =R_{01j0} + R_{0ij}{}^i = -C_j M_{01} + C_i M_j^{\;i} =0.$$
  Hence, since $M_{i}$ is skew-symmetric, we have
  $$C^j R_{0j} = - C^j C_j M_{01} + C_j C_i M^{ij} = - M_{01}\sum_j
  (C_j)^2 = 0.$$
  Hence either $C_j=0$ or $M_{01}=0$.  In the second
  case, $M_{ij}=0$ and Eq. \eqref{eq:mortho} is satisfied.  In both cases, by Eq.
  \eqref{eq:zr2bw-1}
  $$R_{0i01}=R_{0ijk}=0.$$
  Since components of negative weight cannot contribute to an
  invariant,  Lemma \ref{lem:pwzero}  implies that the weight-zero
  components of $R_{abcd}$ also vanish.
\end{proofof}

% \begin{proofof}{Lemma \ref{lem:riczero}}
%   We consider the total skew-symmetrization
%   $$K_{abcd} = R_{[abcd]}=\frac13\, (R_{abcd}+R_{acdb}+R_{adbc}),$$
%   which is non-vanishing by assumption.  We consider the covariant
%   $$Q_{ab} = K_{acde} K_b{}^{cde},$$
%   and use Corollary \ref{cor:pnform} to choose an null-frame so that
%   $$Q_{00} = K_{0ijk} K_0{}^{ijk} = \sum_{ijk} (K_{0ijk})^2 = 0.$$
%   Therefore, $K_{0ijk}=0$, i.e., $K_{abcd}$ has
%   boost-order of zero or less.
%   But Lemma \ref{lem:nzr} implies that $K_{abcd}$ must have strictly
%   negative boost order.  Hence,
%   $$Q_{11} = K_{1ijk} K_1{}^{ijk} = \sum_{ijk} (K_{1ijk})^2 >0,$$
%   and hence by performing a boost as necessary, we have without loss
%   of generality,
%   $$Q_{ab} = \pm \ell_a \ell_b.$$
%   We now proceed as in the proof of Lemma \ref{lem:nzr} to conclude
%   that $R_{abcd}$ has negative boost order.
% \end{proofof}

The {\bf proof of lemma 10} is given in \cite{Mil2004}.

\begin{proofof}{Lemma \ref{lem:zrn}}
Setting
\[ S_{abcd}=R_{(a|ef|b} R_c{}^{ef}{}_{d)},\]
we have
\[ S_{0000} = \sum_{ij} (R_{0i0j})^2.\]
Using Lemma 10, we choose $\ell^a$ such that $S_{0000}=0$,
and hence
\begin{equation}
  \label{eq:Rijvanish}
  R_{0i0j}=0.
\end{equation}
By Corollary \ref{cor:vsiq2},
  \begin{equation}
    \label{eq:zrsumsquares}
    R_{\alpha\beta} = \sum_{\lambda=1}^{p}
    \ovdK{\lambda}{\alpha}\ovdK{\lambda}{\beta} -
    \sum_{\lambda=p+1}^{p+q} \ovdK{\lambda}{\alpha}\ovdK{\lambda}{\beta},
  \end{equation}
  where, without loss of generality, $p<q$, and where the generating
  bivectors $\ovdK{\lambda}{\alpha}$ are linearly independent, null,
  and mutually orthogonal \eqref{eq:mutortho}.  Set
  $$\Xsup{\lambda}{i} =
  \ovdK{\lambda}{0i},\quad \lambda=1,\ldots,p+q,\; i=2,\ldots,n-1,$$
  and note that, by Eq. \eqref{eq:Rijvanish},
  $$R_{0i0j}=\sum_{\lambda=1}^{p} \Xsup{\lambda}{i} \Xsup{\lambda}{j}-
  \sum_{\lambda=p+1}^{p+q} \Xsup{\lambda}{i} \Xsup{\lambda}{j}
  =0,$$
  for all $i,j=2,\ldots, N-1$.  Hence by Corollary
  \ref{cor:indefcoll} we may, without loss of generality, assume that
  \begin{equation}
    \label{eq:normalizedK}
    \ovdK{\lambda}{0i} = \ovdK{\lambda+p}{\!0i},\; \lambda=1,\ldots
    p,\quad\mbox{and}\quad \ovdK{\lambda}{0i}=0,\; \lambda>2p.
  \end{equation}
  For $\lambda=1,\ldots, p$, we set
  $$\ovdE{\lambda}{\alpha} =
  \ovdK{\lambda}{\alpha}+\ovdK{\lambda+p}{\alpha},\quad
  \ovdF{\lambda}{\alpha} =
  \ovdK{\lambda}{\alpha}-\ovdK{\lambda+p}{\alpha};$$
  for $\lambda=p+1,\ldots,q$ we set
  $$\ovdF{\lambda}{\alpha} = \ovdK{\lambda+p}{\alpha}.$$
  Now, Eq. \eqref{eq:zrsumsquares} may be re-expressed as
  \begin{equation}
    \label{eq:zrsumsquares2}
    R_{\alpha\beta} = \sum_{\lambda=1}^{p} 2
    \ovdE{\lambda}{(\alpha}\ovdF{\lambda}{\beta)} +
    \sum_{\lambda=p+1}^{q} \ovdF{\lambda}{\alpha}\ovdF{\lambda}{\beta}.
  \end{equation}
  By Eq. \eqref{eq:normalizedK} the $\ovdF{\lambda}{\alpha},$ are aligned.
  Since they are null and mutually orthogonal, we have by
  Proposition \ref{prop:aligned-ortho} that there exist $F_{ij},
  F_{01}$ and $C_\lambda$ such that
  \begin{equation}
    \label{eq:mortho2}
    F_{ij} F^{ij} - 2 F_{01}{}^2 =0,
  \end{equation}
  and such that
  $$\ovdF{\lambda}{ij} = C_\lambda F_{ij},\quad \ovdF{\lambda}{01} =
  C_\lambda F_{01}.$$
  Setting
  $$E_\alpha = \sum_{\lambda=1}^p C_\lambda \ovdE{\lambda}{\alpha},$$
  we have, by Eq. \eqref{eq:zrsumsquares2},
  $$R_{0i01} =E_{0i} F_{01},\quad R_{0ijk} = E_{0i} F_{jk}.$$

  The assumption $R_{ab}=0$
  implies that
  $$R_{0i} =-R_{0i01} + R_{0ji}{}^j - R_{0i01} = -E_{0i}F_{01} +
  E_{0j} F_i{}^j=0.$$
  However, $F_{ij}$ is skew-symmetric, and hence
  $$E_{0i} R_0{}^i = - E_{0i} E_0{}^i F_{01} + E_{0i} E_{0j} F^{ij} =
  -F_{01} \sum_i (E_{0i})^2   = 0.$$
  Therefore, either $E_{0i}=0$, or $F_{01}=0$.  In the latter case, by Eq. 
  \eqref{eq:mortho} $F_{ij}=0$ as well.  In either case, the
  components of weight 1 necessarily vanish:
  $R_{0i01} = R_{0ijk}=0.$
  Hence, by Lemma \ref{lem:pwzero}, $R_{abcd}$ has negative boost order.
\end{proofof}

\begin{proofof}{Lemma \ref{lem:zr2}}
  We define the following sequence of bivectors
  $$\ovdK{i}{ab} = R_{1iab};$$
  these are aligned because of the assumption of negative boost order.
  Also, by assumption, we have
  $$\suppsub{R}{2}{1i1j} = \ovdK{i}{\alpha} \ovuK{j}{\alpha}=0.$$
  Since $R_{1j}=0$, we can adapt the argument at the end of Lemma
  \ref{lem:nzr} to establish that
  $$R_{1i01}=R_{1ijk}=0.$$
\end{proofof}


\begin{thebibliography}{}


\bibitem{VSI} V. Pravda, A. Pravdov\'a, A.  Coley, R. Milson,  Class. Quantum
  Grav. {\bf 19}, 6213 (2002).
%{\it All spacetimes with vanishing curvature invariants},


\bibitem{kramer}
D. Kramer, H. Stephani, M. MacCallum  and E. Herlt,
{\it Exact Solutions of Einstein's Field Equations}
(Cambridge University Press, Cambridge 1980).



\bibitem{kundt}
W. Kundt,
Z.~Phys. {\bf 163}, 77 (1961).

\bibitem{jordan}
P. Jordan, J. Ehlers and W. Kundt,
Abh. Akad. Wiss. Mainz, Math.-Nat. {\bf 2},
77 (1960).


\bibitem{ss} G. Kofinas, JHEP, {\bf 0108}, 034 (2001); Yu. V. Shtanov and V. Sahni, Int.
J. Mod. Phys. {\bf D11}, 1515
 (2000); {\em ibid.},  Phys. Lett. {\bf B557}, 1 (2001).


\bibitem{wesson} J.M. Overduin and P.S. Wesson, Phys. Reports.
{\bf 283}, 303
 (1997).

\bibitem{cmpppz}
A. Coley,  R. Milson,  N. Pelavas,  V. Pravda, A. Pravdov\'a,  and R. Zalaletdinov,
Phys. Rev. D {\bf 67}, 104020 (2003).
%{\it Generalizations of pp-wave spacetimes in higher dimensions}, 

\bibitem{Algclass} R. Milson, A. Coley, V. Pravda, A. Pravdov\'a,
{\it Aligned null directions and tensor classification}, Int. J. Geom. Meth. Mod.
Phys. [gr-qc/0401010] (2004).

\bibitem{newcqg}
A. Coley,  R. Milson,  V. Pravda,  and A. Pravdov\'a, Class. Quantum Grav. {\bf 21},  L35 (2004).
%{\it Classification of the Weyl tensor in higher dimensions}, 


\bibitem{Bianchi} V. Pravda, A. Pravdov\'a, A. Coley, R. Milson, 
Class. Quantum Grav. {\bf 21}, 2873 (2004).%gr-qc/0401013 (2004).
%{\it Bianchi identities in higher dimensions},

\bibitem{penrind} R. Penrose and W. Rindler, \emph{Spinors and
    Space-time}, Volumes 1 \& 2 (Cambridge University Press, Cambridge 1986).



\bibitem{petrov} A. Petrov, \emph{Einstein Spaces} (Pergamon Press, Oxford 1969).


\bibitem{BiPr} J. Bi\v c\' ak and V. Pravda,
Class. Quantum Grav. {\bf 15}, 1539 (1998).
%{\it Curvature invariants in type N spacetimes},

\bibitem{Pravda} V. Pravda,
Class.  Quantum Grav. {\bf 16}, 3321 (1999).
%{\it Curvature invariants in type III spacetimes},

\bibitem{amati}
D. Amati and C. Klim\v c\'\i k,
Phys. Lett. {\bf B 219}, 443 (1989);
D. Amati and C. Klim\v c\'\i k, Phys. Lett. {\bf B210}, 92 (1988).

\bibitem{HS}
G.T. Horowitz and A.R. Steif, Phys. Rev. Lett.
{\bf 64} (1990) 260: {\em ibid.}  Phys.\ Rev.\ D {\bf
42}, 1950 (1990).


\bibitem{coley} A.A. Coley,
Phys. Rev. Letts,
{\bf 89}, 281601
(2002).



\bibitem{matsaev}
R.R. Metsaev, Nucl. Phys. {\bf B625}, 70 (2002);
R.~R.~Metsaev and A.~A.~Tseytlin,
Phys. Rev.  D
{\bf 65}, 126004
(2002).


\bibitem{tseytlin}
G. T.Horowitz  and A. A. Tseytlin
Phys. Rev.  D
{\bf 51}, 2896
(1995).


\bibitem{rutone} J.~G.~Russo and A.~A.~Tseytlin, Nucl.
Phys. B {\bf 448}, 293 (1995);
 A.~A.~Tseytlin, Class. Quant. Grav. {\bf 12}, 2365 (1995).



\bibitem{RT} J.~G.~Russo and A.~A.~Tseytlin,
JHEP {\bf 0204}, 021 (2002); M. Blau, J. Figueroa-O'Farrill, C. Hull and G. Papadopoulos
JHEP {\bf 0201},  047 (2002);
{\em ibid.}
Class. Quant. Grav. {\bf19}, L87 (2002);
P. Meessen, Phys. Rev.  D {\bf 65}, 087501
(2002); A.~A.~Tseytlin,
Phys. Rev. D {\bf 47}, 3421 (1993); A.~A.~Tseytlin,
Nucl. Phys. B {\bf 390}, 153 (1993).



\bibitem{russot} J. G. Russo and A.A. Tseytlin, JHEP
{\bf 0209}, 035 (2002).



\bibitem{GuevenAD}
R.~Gueven,
Phys. Lett. B {\bf 191}, 275 (1987);
 J. Kowalski-Glikman,  Phys. Lett.  {\bf B150}, 125
(1985).

\bibitem{brane}
V. Rubakov and M.  Shaposhnikov
Phys. Lett.  {\bf B125}, 139 (1983);
 N.
Arkani-Hamed, S. Dimopoulos and G. Dvali
Phys. Lett.  {\bf B429}  263 (1998); L.
Randall and R.  Sundrum
Phys. Rev. Letts. {\bf 83} 3370 \& 4690
(1999).

\bibitem{mam} J.~Maldacena and L.~Maoz,
JHEP
{\bf 0212}, 046 (2002).

\bibitem{GSW} M.~B.~Green, J.~H.~Schwarz and E.~Witten,
\emph{Superstring Theory. Vol. 1: Introduction}
(Cambridge University Press, Cambridge 1987).

\bibitem{MilSPT} G. Bergqvist and J. M. M. Senovilla, 
Class. Quantum Grav. {\bf 18},  5299 (2001).



\bibitem{stewart}
J. Stewart, \emph{Advanced general relativity} (Cambridge University
Press, Cambridge 1990).

\bibitem{fulhar} W.~Fulton and J.~Harris, \emph{Representation theory}
(Springer-Verlag, New York 1991).

\bibitem{Mil2004} R. Milson, preprint: {\it Aligned null directions for higher dimensional Weyl tensors}.


%%\bibitem{NPform}
%%E. T. Newman, R. Penrose, {\it An approach to gravitational radiation by a method of spin coeff
%%icients}, J. Math. Phys. {\bf 3}, 566 (1962).



\end{thebibliography}
\end{document}